%% file: SP_overview_arxiv.tex
\documentclass[journal]{IEEEtran}

\usepackage{cite}

\usepackage{amsmath}
\usepackage{amssymb}
\usepackage{amsfonts}
\usepackage{graphicx} 
\usepackage{subfigure}

\input{shortcuts}

\usepackage{tikz}
\usetikzlibrary{decorations}
\usetikzlibrary{decorations.pathmorphing}
\usetikzlibrary{decorations.text}
\usetikzlibrary{positioning}
\usetikzlibrary{shapes.geometric}
\usetikzlibrary{patterns}
\usetikzlibrary{decorations.pathreplacing}
\usetikzlibrary{shapes,backgrounds}

\title{Efficient DSP and Circuit Architectures for Massive MIMO: State-of-the-Art and Future Directions}

\author{Liesbet Van der Perre, Liang Liu and Erik
  G. Larsson\thanks{L. Van der Perre is with KU Leuven, Belgium and
    Lund University, Sweden. Email:
    \texttt{liesbet.vanderperre@kuleuven.be}}\thanks{L. Liu is with
    Lund University, Sweden.}\thanks{E. G. Larsson is with Link\"oping
    University, Dept. of Electrical Engineering (ISY), 581 83
    Link\"oping, Sweden. The work of E. G. Larsson was supported in
    part by the Swedish Research Council (VR) and ELLIIT.} \\
Overview paper (invited)} \date{}

\begin{document}
\maketitle

\begin{abstract}
Massive MIMO is a compelling wireless access concept that relies on
the use of an excess number of base-station antennas, relative to the
number of active terminals. This technology is a main component of 5G
New Radio (NR) and addresses all important requirements of future
wireless standards: a great capacity increase, the support of many
simultaneous users, and improvement in energy efficiency.

Massive MIMO requires the simultaneous processing of signals from many
antenna chains, and computational operations on large matrices. The
complexity of the digital processing has been viewed as a fundamental
obstacle to the feasibility of Massive MIMO in the past. Recent
advances on system-algorithm-hardware co-design have led to extremely
energy-efficient implementations. These exploit opportunities in
deeply-scaled silicon technologies and perform partly distributed
processing to cope with the bottlenecks encountered in the
interconnection of many signals. For example, prototype ASIC
implementations have demonstrated zero-forcing precoding in real time
at a 55 mW power consumption (20 MHz bandwidth, 128 antennas,
multiplexing of 8 terminals).  Coarse and even error-prone digital
processing in the antenna paths permits a reduction of consumption
with a factor of 2~to~5.  This article summarizes the fundamental
technical contributions to efficient digital signal processing for
Massive MIMO. The opportunities and constraints on operating on low-complexity RF and analog hardware chains are clarified. It illustrates
how terminals can benefit from improved
energy efficiency. The status of technology and real-life prototypes
discussed. Open challenges and directions for future research are
suggested.
\end{abstract}

\input{introduction.tex}

The rest of the paper is organized as follows. First, basic concepts
and notation are introduced. Next, we provide a complexity analysis
considering computation as well as data transfer. The following
section zooms in on the RF and front-end, highlighting the
opportunities and constraints of relaxing their specifications in the
large-number-of-antennas regime. Subsequently, the central detector and precoder
blocks are detailed and major complexity reductions facilitated by
algorithm-hardware co-design are demonstrated. Signal processing
leveraging on error-resilient circuits in the per-antenna
functionality is discussed next, and
consequent energy savings are illustrated.  Further we introduce
introduces the increased reliability that can be delivered on
complexity terminals.  Finally, in the conclusions we discuss validation performed in
real-life test beds, summarize opportunities and constraints in
efficient processing for Massive MIMO systems, and suggest future
research directions.

\input{mamimo-concept.tex}

\section{Signal Processing and Data Transfer Complexity Assessment}

\input{complexity.tex}

\section{Analog and RF Processing: Relax with Caution!}\label{sec:RF}

\input{RFrelax.tex}

\section{Algorithm-Hardware Co-Design for   Precoding and Detection}
\label{sec:precoding-decoding}

\input{precode_decode.tex}

\section{Per-Antenna Chain processing at the Semiconductor Edge}
\label{sec:per-antenna}

\input{PAntP.tex}

\section{Terminals: Increased Reliability with Low-Complexity Signal Processing}
\label{sec:terminal}

\input{terminal.tex}

\section{Demonstrations, Conclusions and Future Directions}\label{sec:conclusions}

\input{conclusions.tex}

\section{Acknowledgment}
The authors thank their colleagues and collaborators in the European FP7-MAMMOET
project for the nice cooperation that truly progressed Massive MIMO technology.

\input{SP_overview_final.bbl}

\clearpage

\textbf{Liesbet Van der Perre} is Professor at the Department of
Electrical Engineering at the KU Leuven in Leuven, Belgium and a guest
Professor at the Electrical and Information Technology Department at
Lund University, Sweden. Dr. Van der Perre was with the
nano-electronics research institute imec in Belgium from 1997 till
2015, where she took up responsibilities as senior researcher, system
architect, project leader and program director. She was appointed
honorary doctor at Lund University, Sweden, in 2015. She was a
part-time Professor at the University of Antwerp, Belgium, from 1998
till 2002.  She received her Ph.D. degree from the KU Leuven, Belgium,
in 1997.

Her main research interests are in wireless communication, with a
focus on physical layer and energy efficiency in transmission,
implementation, and operation.  Prof. L. Van der Perre was the
scientific leader of FP7-MAMMOET, Europe's prime project on Massive
MIMO technology. Dr. Van der Perre has been serving as a scientific
and technological advisor, reviewer and jury member for companies,
institutes, and funding agencies. She is a member of the Board of
Directors of the company Zenitel since 2015.  Liesbet Van der Perre is
an author and co-author of over 300 scientific publications. She was a
system architect for the OFDM ASICs listed in the \emph{IEEE
  International Solid State Circuit Conference (ISSCC's) Best of 50
  Years} papers in 2003. She co-authored the paper winning the
DAC/ISSCC 2006 design contest.

~

\textbf{Liang Liu} is an Associate Professor at Electrical and
Information Technology Department, Lund University, Sweden. He
received his Ph.D. in 2010 from Fudan University in China.  In 2010,
he was with Electrical, Computer and Systems Engineering Department,
Rensselaer Polytechnic Institute, USA as a visiting researcher. He
joined Lund University as a Post-doc researcher in 2010 and is now
associate professor there.  His research interest includes signal
processing for wireless communication and digital integrated circuits
design.  Liang is active in several EU and Swedish national projects,
including FP7 MAMMOET, VINNOVA SoS, and SSF HiPEC, DARE.  He is a
board member of the IEEE Swedish Solid-State Circuits/Circuits and
Systems chapter.  He is also a member of the technical committees of
VLSI systems and applications and CAS for communications of the IEEE
Circuit and Systems society.

~

\textbf{Erik G. Larsson} received the Ph.D. degree from Uppsala
University, Uppsala, Sweden, in 2002.

He is currently Professor of Communication Systems at Link\"oping
University (LiU) in Link\"oping, Sweden. He was with the KTH Royal
Institute of Technology in Stockholm, Sweden, the George Washington
University, USA, the University of Florida, USA, and Ericsson
Research, Sweden.  In 2015 he was a Visiting Fellow at Princeton
University, USA, for four months.  His main professional interests are
within the areas of wireless communications and signal processing. He
has co-authored some 150 journal papers on these topics, he is
co-author of the two Cambridge University Press textbooks
\emph{Space-Time Block Coding for Wireless Communications} (2003) and
\emph{Fundamentals of Massive MIMO} (2016). He is co-inventor on 18
issued and many pending patents on wireless technology.

He is a member of the IEEE Signal Processing Society Awards Board
during 2017--2019.  He is an editorial board member of the
\emph{IEEE Signal Processing Magazine} during 2018--2020. 
From 2015 to 2016 he served as chair of the IEEE
Signal Processing Society SPCOM technical committee.  From 2014 to
2015 he was chair of the steering committee for the \emph{IEEE
  Wireless Communications Letters}.  He was the General Chair of the
Asilomar Conference on Signals, Systems and Computers in 2015, and its
Technical Chair in 2012.  He was Associate Editor for, among others,
the \emph{IEEE Transactions on Communications} (2010-2014) and the
\emph{IEEE Transactions on Signal Processing} (2006-2010).
  
He received the IEEE Signal Processing Magazine Best Column Award
twice, in 2012 and 2014, the IEEE ComSoc Stephen O. Rice Prize in
Communications Theory in 2015, the IEEE ComSoc Leonard G. Abraham
Prize in 2017, and the IEEE ComSoc Best Tutorial Paper Award in 2018. 
He is a Fellow of the IEEE.

\end{document}

%% file: shortcuts.tex
\def\ba{{\boldsymbol{a}}}

\def\bB{{\boldsymbol{B}}}

\def\bd{{\boldsymbol{d}}}

\def\bg{{\boldsymbol{g}}}
\def\bG{{\boldsymbol{G}}}

\def\bI{{\boldsymbol{I}}}

\def\bL{{\boldsymbol{L}}}

\def\bQ{{\boldsymbol{Q}}}

\def\bR{{\boldsymbol{R}}}

\def\bs{{\boldsymbol{s}}}

\def\bT{{\boldsymbol{T}}}

\def\bw{{\boldsymbol{w}}}

\def\bx{{\boldsymbol{x}}}
\def\bX{{\boldsymbol{X}}}

\def\by{{\boldsymbol{y}}}

\def\bZ{{\boldsymbol{Z}}}

\newcommand{\pp}[1]{{\left( #1 \right)}}

\newcommand*\circled[1]{\tikz[baseline=(char.base)]{
            \node[shape=circle,draw,inner sep=0pt] (char) {#1};}}

%% file: introduction.tex
\section{Introduction}

\begin{figure}[t!]
\centering
\includegraphics[width=0.45\textwidth]{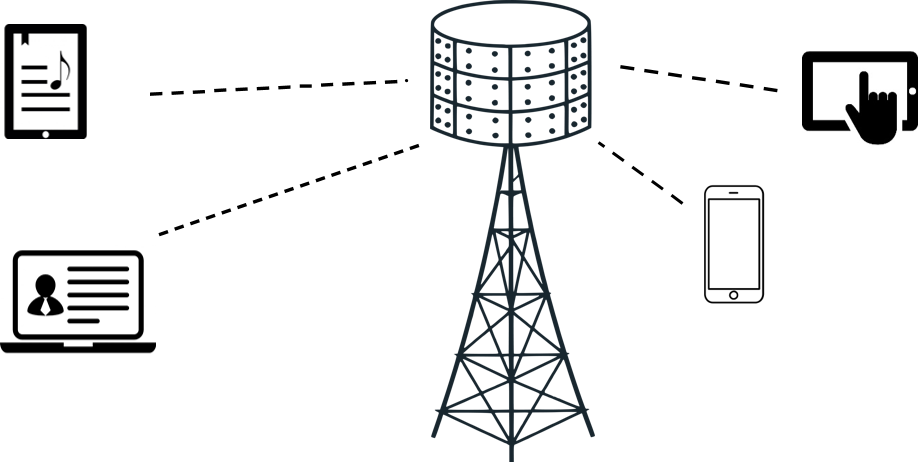}
  \caption{Massive MIMO exploits large antenna arrays at the base stations, to spatially multiplex many terminals.\label{fig:systemPict}}
\end{figure}

Massive MIMO is an efficient sub-6 GHz physical-layer technology for
wireless access, and a key component of the 5G New Radio (NR)
interface \cite{nr}. The main concept is to use large antenna arrays
at base stations to simultaneously serve many autonomous terminals, as
illustrated in Figure~\ref{fig:systemPict}
\cite{Larsson2014,Marzetta16book}.  Smart processing at the
array exploits differences among the propagation signatures of the
terminals to perform spatial multiplexing. Massive MIMO offers two
main benefits:
\begin{enumerate}
\item Excellent spectral efficiency, achieved by spatial multiplexing
  of many terminals in the same time-frequency resource  \cite{Harris2017JSAC,SEU_NUPT}. Efficient multiplexing requires
  channels to different terminals to be sufficiently distinct.  Theory
  as well as experiments have demonstrated that this can be achieved
  both in line-of-sight and in rich scattering.

\item Superior energy efficiency, by virtue of the array gain, that
  permits a reduction of radiated power. Moreover, the ability to
  achieve excellent performance while operating with low-accuracy
  signals and linear signal processing further enables considerable
  savings in the power required for signal processing.
\end{enumerate}

This overview paper focuses on sub-6 GHz Massive MIMO systems
implemented with fully digital per-antenna signal processing.  Massive
MIMO at mmWave frequencies is also possible, and can benefit from the
large bandwidth available at these frequencies.  Propagation and
hardware implementation aspects are different at mmWaves; for example,
hybrid analog-digital beamforming approaches are typically considered
\cite{Molish2017}. However, this is not discussed further
here.

The complexity of the signal processing has been considered a
potential obstacle to actual deployment of Massive MIMO technology. An
obvious concern is how operations on large matrices and the
interconnection of the many antenna signals can be efficiently
performed in real-time. Moreover, real-life experiments have shown
that the channel responses to different terminals can be highly
correlated in some propagation environments. Appropriate digital
signal processing hence needs to feature interference suppression
capabilities, which further increases complexity.

\begin{figure}[t!]
\centering
\includegraphics[width=0.4\textwidth]{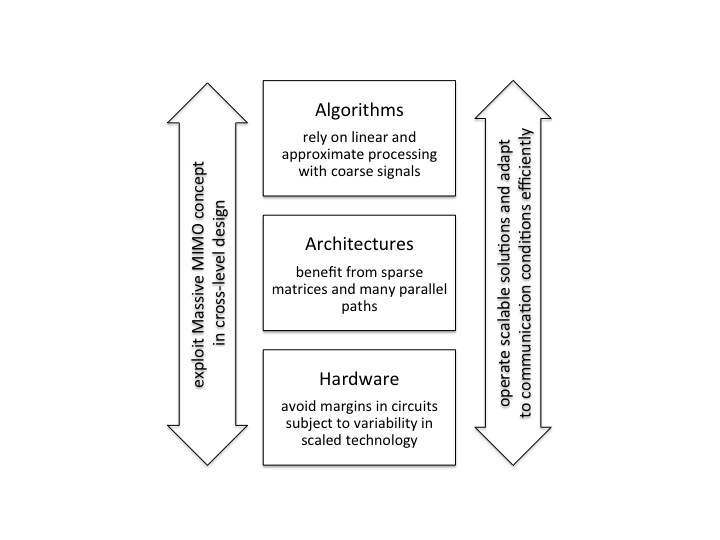}
  \caption{Massive MIMO opens up new hardware-software co-design opportunities for low-complexity circuitry.\label{fig:cross_level}}
\end{figure}

This paper discusses the digital signal processing required to realize
the Massive MIMO system concept, and examines in detail the co-design
of algorithms, hardware architecture, and circuits
(Figure~\ref{fig:cross_level}).  Unconventional, low-complexity
digital circuitry implementations in deeply scaled silicon are
possible, despite (and thanks to) the excess number of antenna
signals. A careful choice of algorithmic and circuit parameters
permits considerable reduction of the average energy consumption.
Terminals in turn can be implemented at low complexity while
benefiting from the channel hardening effect, that offers increased
reliability.

Proof of concept implementations and demonstrations have revealed
constraints that turned out more harsh than anticipated in initial
theoretical assessments. This concerns the interconnection of the
signals from all antennas, which poses a bottleneck that partly
necessitates distributed processing.  Also, relaxing the specifications
of the analog and RF chains can result in higher distortion both in-band and out-of-band than initially anticipated, as hardware imperfections can in general not be considered uncorrelated.
 
%% file: mamimo-concept.tex
\section{Massive MIMO System Model} \label{sec:concept}

This section introduces the notation for MIMO transmission that is
used in the paper. Further details can be found in, for example,
\cite{Marzetta16book}. We consider the block-fading model where the
time-frequency domain is partitioned into coherence intervals within
which the channel is static. The number of samples in each coherence
interval is equal to the coherence time in seconds multiplied by the
coherence bandwidth in Hertz.  For the signal processing algorithms
discussed in this paper, it does not matter whether there is coding
across coherence intervals or not.

In every coherence interval, a flat fading complex baseband channel
model applies.  Let $M$ be the number of antennas at the base station,
and $K$ the number of terminals served simultaneously. Also, denote by
$\bg_k$ the $M$-vector of channel responses between the $k$th terminal
and the array.  Then on uplink, for every sample in the coherence interval,
\begin{align}\label{eq:ul}
\by = \sum_{k=1}^K \bg_k x_k + \bw,
\end{align}
where $\by$ is an $M$-vector comprising samples received at the base
station array, $\{x_k\}$ are symbols sent by the $k$th terminal, and
$\bw$ is noise.  On downlink, assuming linear precoding,
\begin{align}\label{eq:dl}
y_k = \bg_k^T \sum_{k'=1}^K \ba_{k'}  x_{k'} + w_k.
\end{align}
where $y_k$ is the sample received by the $k$th terminal, $\ba_k$ is a
precoding vector associated with the $k$th terminal, $x_k$ is the
symbol destined to the $k$th terminal, and $w_k$ is $CN(0,1)$ receiver
noise.

The base station forms a channel estimate, $\hat\bg_k$, of $\bg_k$ for
each terminal $k$ by measurements on uplink pilots.  Channel
estimation is discussed extensively in for example
\cite{Marzetta16book} (for independent Rayleigh fading) and
\cite{bjornson2017massive} (for correlated fading models).

On uplink, the data streams from the terminals are detected through
linear processing. This entails multiplication of $\by$ with a vector,
$\ba_k$ for each terminal, yielding the scalar $ \ba_k^H \by$.  Common
choices of the detection vector $\ba_k$ include
\begin{align}
\begin{cases}
\text{max.-ratio:} & \ba_k = \alpha_k \hat\bg_k \\
\text{zero-forcing:} & \ba_k = \alpha_k \left[ \hat\bG  (\hat\bG^H\hat\bG)^{-1}  \right]_{(:,k)} \\
\text{MMSE:} & \ba_k = \alpha_k  \left[ \hat\bG (\hat\bG^H\hat\bG + \bI)^{-1}  \right]_{(:,k)}
\end{cases}
\end{align}
where $\alpha_k$ is a normalizing constant (different for the three
methods), and $\hat\bG=[\hat\bg_1,\ldots,\hat\bg_K]$. The result of
this linear processing will comprise the desired signal, embedded in
additive interference and noise.

On downlink, channel reciprocity is leveraged. Low-complexity
front-ends typically introduce non-reciprocity 
and
this non-reciprocity needs to be compensated for; see
Section~\ref{sec:RF}.  The base station forms the transmitted vector
$\sum_k \ba_kx_k$ in (\ref{eq:dl}) where the precoding vector $\ba_k$
is given by:
\begin{align}
\begin{cases}
\text{max.-ratio:} & \ba_k = \alpha_k \hat\bg_k^* \\
\text{zero-forcing:} & \ba_k = \alpha_k \left[ \hat\bG^*(\hat\bG^T\hat\bG^*)^{-1}  \right]_{(:,k)} \\
\text{regularized zero-forcing:} & \ba_k = \alpha_k  \left[ \hat\bG^*(\hat\bG^T\hat\bG^* + \lambda \bI)^{-1}  \right]_{(:,k)}
\end{cases}
\end{align}
where, again, $\{\alpha_k\}$ are normalizing constants and $\lambda$
is a regularization parameter. 
The  signal received at the terminal will contain the symbol of interest, plus
additive interference and noise.
 
Many variations are possible and detection and precoding that take
multi-cell interference into account are also possible
\cite{bjornson2017massive,li2017massive}.

%% file: complexity.tex
Both Massive MIMO base stations and terminals can be implemented with
significantly better energy efficiency compared to in conventional
systems. This is possible owing to a combination of effects. First,
the array gain permits a reduction of the radiated power. Second, the
large number of constituent signals promotes excellent performance
while operating relatively simple algorithms on coarse signals.

In this section we focus on the processing at the base station
side. The opportunity to reduce terminal-side complexity is discussed
in Section~\ref{sec:terminal}. First, a high-level assessment of the
signal processing requirements, in terms of number of computations, is
presented.  The data transfer and interconnection of signals poses a
distinct bottleneck.  Hence, next a distributed processing approach is
presented to balance performance and complexity.

\subsection{Computational Complexity}

\begin{figure*}[t!]
\centering
\includegraphics[width=0.8\textwidth]{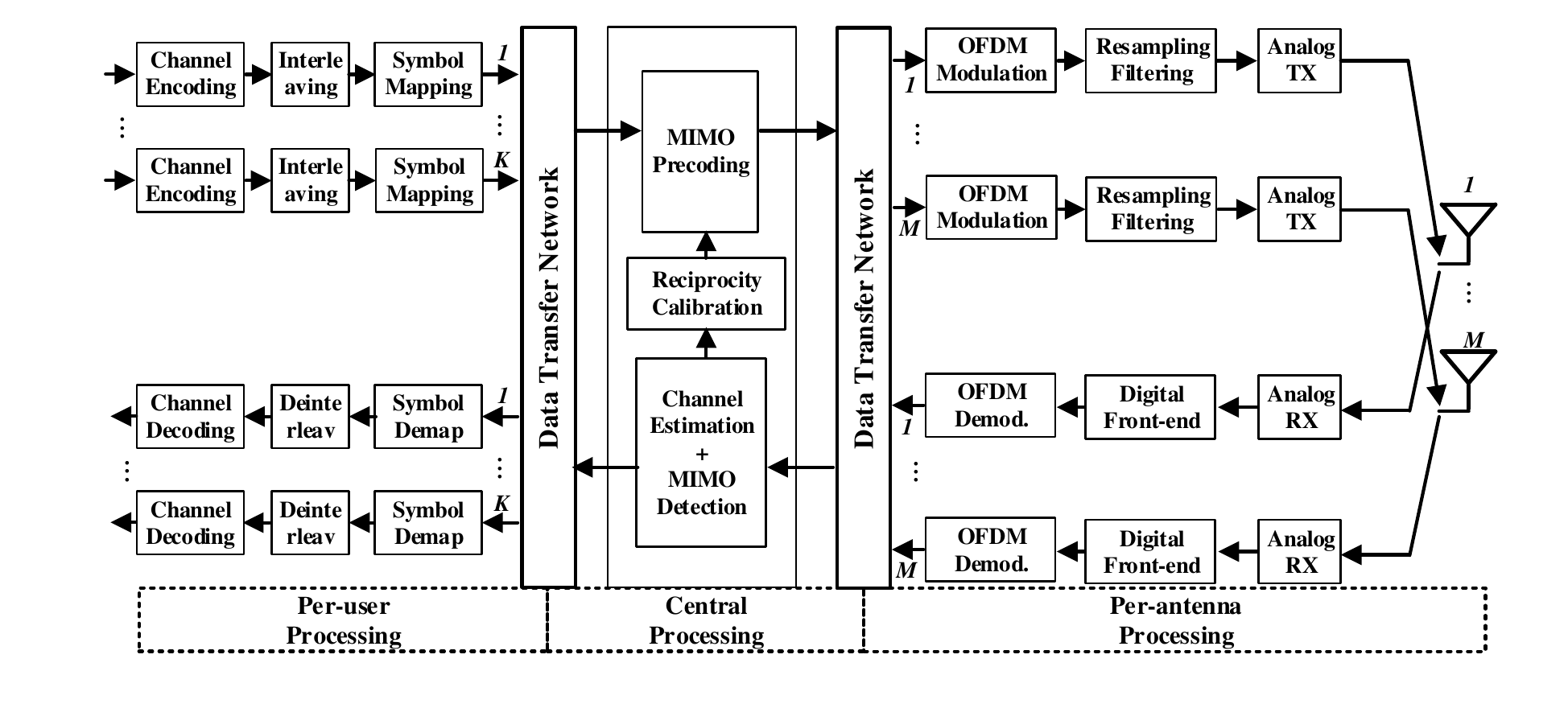}
\caption{Signal processing in an OFDM-based massive MIMO system for $M$ BS antennas and $K$ UEs.}
\label{fig:proc_dist}
\end{figure*}

We first analyze the computational complexity of a Massive MIMO base
station. Figure~\ref{fig:proc_dist} shows a high-level block diagram
of the signal processing for an OFDM-based massive MIMO system. Other
modulation options can be used, and single-carrier schemes may be
preferred. 
The overall partition
of the processing presented here will still hold.

The processing in Massive MIMO systems is logically grouped into three
categories:
\begin{enumerate}
	\item The outer modem performing symbol (de)mapping,
	(de)interleaving and channel (de)coding. This processing
	performed on the transmit/receive bits applies to each User
	Equipment (UE) individually.

\item The inner modem comprising channel estimation, and detection and precoding of the
	uplink and downlink data, respectively. This central
	processing aggregates/distributes data from/to all the antenna
	chains.

\item The per-antenna processing which primarily consists of the analog and digital front-end (mainly re-sampling and filtering) and OFDM processing.
\end{enumerate}

We identify inherent parallelism and observe that the processing
complexity scales with the number of BS antennas, $M$, the number of
UEs, $K$, or both \cite{Steffen2016SIPS}:
\begin{itemize}
	\item Per-antenna processing: Scales with $M$ as each antenna
	requires OFDM (de)modulation and a digital/analog front-end.
        
	\item Central processing: Scales with $M$ and $K$.
        
	\item Per-user processing: Scales with $K$.
\end{itemize}

The number of digital signal processing operations performed in the
sub-systems provides a high-level estimate of
complexity. Table~\ref{tb:complexity_case} gives numbers for a sample
system with $M=100$ antennas at the base-station and $K=10$
simultaneous terminals. It is acknowledged that these estimates
represent an over-simplification, as the nature and precision of the
operations will be an important determining factor in the eventual
hardware complexity and power consumption.

\begin{table}[t!]
	\caption{Estimated number of DSP operations in GOPS, for $M=100$ and $K=10$, $20$ MHz bandwidth,
        and $3$ bps/Hz (16-QAM, code rate 3/4).\label{tb:complexity_case}}
	 \centering
	\begin{tabular}{ | c | c | c | c | }
		\hline
		\textbf{\textit{Subcomponent}}  & \textbf{\textit{Downlink data (DL)}} & \textbf{\textit{Uplink data (UL)}}  & \textbf{\textit{Training}} \\
		& \textbf{\textit{[GOPS]}}  & \textbf{\textit{[GOPS]}} & \textbf{\textit{[GOPS]}} 	\\\hline
		Inner modem &	175	& 520 &	290 \\\hline
		Outer modem &	7	& 40  &	0 	\\\hline
		Per-antenna DSP &  920	& 920 &	920 \\\hline
	\end{tabular}
\end{table}

Table~\ref{tb:complexity_case} demonstrates that the collective
per-antenna digital processing is demanding, and requires a
minimal-complexity implementation.  Interestingly, the per-antenna
processing does not need to be performed with high precision to offer
very good performance. An in-depth analysis and efficient
implementation options are presented in Section~\ref{sec:per-antenna}.

For the inner modem processing in Massive MIMO, a high degree of
reconfigurability is desired in order to adapt to changing operating
conditions, such as the number of connected UEs, and their SNRs/path
losses. Section~\ref{sec:precoding-decoding} discusses efficient
algorithm-hardware co-design solutions for the Massive MIMO precoding
and detection.

Furthermore, reciprocity calibration needs to be performed
occasionally. Elegant solutions have been proposed and demonstrated,
see Section~\ref{sec:RF}. 

Channel coding clearly is an essential component
of the wireless transmission, yet it is not Massive MIMO-specific and
therefore not further treated in this paper.

\subsection{Signal Interconnection and Data Transfer Complexity}

The transfer of data between processing components creates a
significant challenge, as the amount of signals and data to be
aggregated/distributed from/to all the antennas is very high. The
required data shuffling rate between the per-antenna processing and
the central processing is \cite{Steffen2016SIPS}
\begin{equation}
R_\textnormal{antennas2central}=M\times R_\textnormal{OFDM}\times W,
\label{eq:data_size}
\end{equation}
where $R_\textnormal{OFDM}$ is the sampling rate after OFDM processing
and $W$ is the word-length of one data sample. For a 100-antenna
20~MHz bandwidth system, the sampling rate $R_\textnormal{samp}$ at
each antenna is 30.72 $MS/s$ and thus
\begin{equation}
R_\textnormal{OFDM}=R_\textnormal{samp} \times \frac{N_\textnormal{data}}{N_\textnormal{sub}+N_\textnormal{CP}}=16.8\
\mbox{MS/s},
\end{equation}
where $N_\textnormal{data}$, $N_\textnormal{sub}$, and
$N_\textnormal{CP}$ are the number of data subcarriers, the total
number of subcarriers, respectively the number of cyclic prefix
samples.  Assuming that 24 bits are used for one complex sample,
$R_\textnormal{antennas2central}$ equals 40.32 $Gb/s$. This
requirement is an order of magnitude higher than in a conventional
system.

\begin{figure}[t!]
	\centering \includegraphics[width=0.45\textwidth]{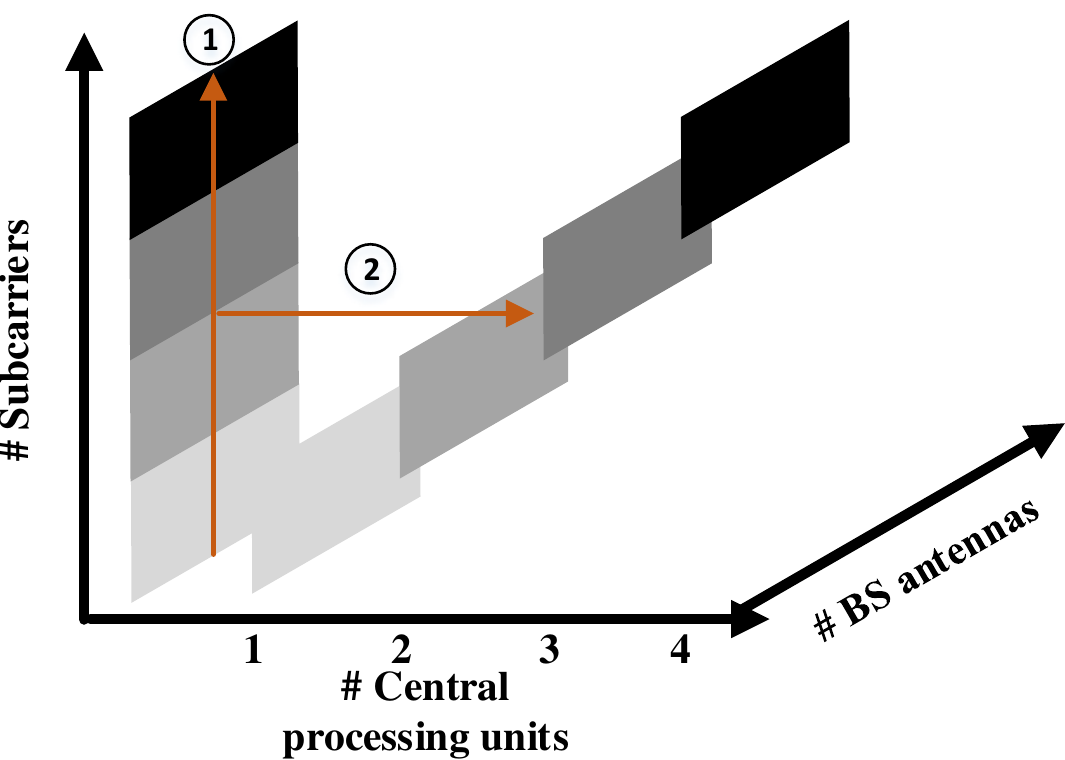} \caption{Illustration
	of the data shuffling between the per-antenna and central
	processing.}  \label{fig:data_network}
\end{figure}

Additionally, the data transfer network must re-organize data among
different dimensions. Figure~\ref{fig:data_network} illustrates the
uplink data shuffling between the per-antenna and the central
processing. First, \circled{1} in the figure, the data shuffling
network aggregates data samples of all subcarriers from all antenna
chains. Next, \circled{2} in the figure, it divides the entire data
into bandwidth chunks depending on the number of central processing
units in the system, and distributes the data to the corresponding
processing unit.

This high data transfer requirements has motivated the development of
decentralized processing architectures, which are introduced next.

\subsection{Decentralized Processing}

Depending on the selected MIMO processing algorithms, both the
processing performed in the per-antenna and in the central units, and
the communication between these two, will influence the resulting
system performance and overall complexity. For instance, the
maximum-ratio precoding operation $\sum_k \alpha_k \hat\bg_k^*x_k$ can
be performed in each antenna path in a distributed manner, whereas
the zero-forcing algorithm requires centralized processing, specifically for the
inversion of the Gram matrix $(\hat\bG^H\hat\bG)^{-1}$.

\begin{figure}[t!]
\centering
\includegraphics[width=0.45\textwidth]{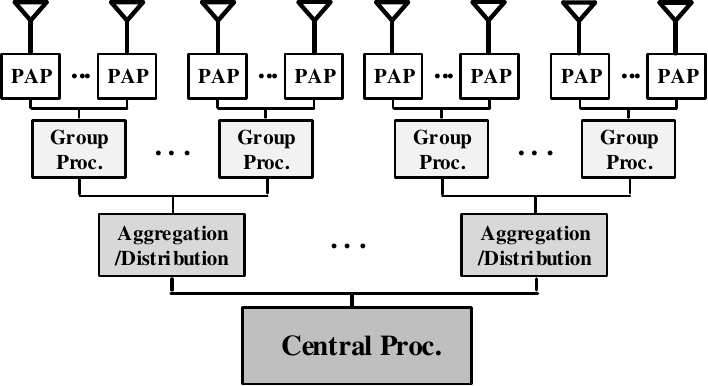}
\caption{Decentralized processing architecture, performing group-based operations
between the per-antenna processing (PAP) and the central unit.}
\label{fig:decentralized_processing}
\end{figure}

Decentralized processing enables parallel computing and offers a
 balanced trade-off between system performance and data transfer
 requirements \cite{SYA2012,DBLP:journals/pieee/PuglielliTLMLTW16,li2017decentralized,DBLP:conf/sigcomm/YangLYFTHZZ13}. The
 authors of \cite{li2017decentralized} propose a decentralized
  architecture for both uplink and downlink, illustrated
 in Figure~\ref{fig:decentralized_processing}. Instead of aggregating
 the full channel state information and transmit/received data vectors
 at the centralized processing node, $M$ antenna nodes are grouped
 into $B$ equally sized groups, each serving $C$ antenna nodes. A middle
-level processing node, labeled group processor, is introduced between
 the per-antenna and central processor to handle the corresponding
 data dedicated to the group of $C$ antenna nodes. As a result, a
 limited amount of data is then aggregated/distributed to/from the
 central processor, relaxing the requirements on the data transfer
 network. For instance, the Gram matrix calculation
 $\bZ=\hat{\bG}^H\hat{\bG}$ can be rewritten as
\begin{equation}
\bZ = \sum_{b=1}^{B}\hat{\bG}_b^H\hat{\bG}_b,
\label{eq:dec_proc}
\end{equation}
where $\hat{\bG}_b \in \mathbb{C}^{C \times K}$ is the local channel
estimate for each group of $C$ antennas. The decentralized processing
is performed such that the terms $\hat{\bG}_b^H\hat{\bG}_b$ are
computed at each group processor locally, and the results are
aggregated at the central processor for the final summation. The
tree-like distributed processing architecture is further elaborated
in \cite{Bertilsson2017}, with special focus on modularity and
scalability. Especially, the trade-off between data processing,
storage, and shuffling is investigated for maximum-ratio transmission,
zero-forcing, and MMSE algorithms.

%% file: RFrelax.tex
In traditional base stations, the RF electronics and analog
front-ends, and the power amplifiers specifically, consume most of the
power \cite{Auer2011}. In Massive MIMO, thanks to the array gain
provided by the closed-loop beamforming, much less radiated power is
needed for the data transmission.  This facilitates a significant
reduction of the RF complexity and power consumption compared to
conventional systems.

The hardware in any wireless transceiver will introduce distortion,
and the most important source of distortion is nonlinearities in power
amplifiers and quantization noise in A/D-converters.  A commonly used
model in the literature has been that this distortion is additive and
uncorrelated among the
antennas \cite{bjornson2017massive}.
If this were the case, then the effects of hardware imperfections
would average out as the number of antennas is increased, in a similar
way as the effects of thermal noise average out. In more detail,
consider the linear processing in the uplink, $\ba_k^H\by$; see
Section~\ref{sec:concept}. The essence of the argument is that if the
received signal at the array, $\by$, is affected by uncorrelated
additive distortion noise $\bd$, then the effective power of the
useful signal after beamforming processing, $\ba_k^H\bg_k$, would grow
as $M$ whereas the power of the distortion, $\ba_k^H\bd$, would be
constant with respect to $M$
(see \cite{bjornson2017massive} for
more precise analyses).  But unfortunately, this model does not
accurately describe the true nature of the hardware distortion.

To understand why, fundamentally, the distortion is correlated among
the antennas, consider the downlink in the special case of a single
terminal in line-of-sight. Then the signal radiated by the $m$th
antenna is simply a phase-shifted version of the signal radiated by
the first antenna ($m=1$). The distortion arising from an amplifier
nonlinearity at the $m$th antenna is phase-shifted by the same amount
as the signal. Hence, if all amplifiers have identical characteristics
(a weak assumption in practice), the distortion is beamformed into the
same direction as the signal of interest, and receives the same array
gain as that signal of interest. That is, the effects of the
distortion do scale proportionally to $M$ rather than disappearing as
$M$ is increased.  In this case, the covariance matrix of the
distortion, when viewed as an $M$-vector $\bd$, has rank one.  A
similar effect exists on the uplink, when the nonlinearities in
low-noise amplifiers are
considered \cite{DBLP:journals/corr/abs-1712-09612}.

In the remainder of this section, we discuss the specifics of
distortion arising from amplifier nonlinearities and finite-resolution
A/D-converters in more detail. We furthermore discuss the impact and
calibration of RF front-end non-reciprocity.

\subsection{Power Amplifiers Benefit from the Large Array}

The required output power of a Massive MIMO base station can be
reduced inversely proportionally to the square root of number of BS
antennas, or even linearly in operating regimes with good channel
estimation quality, thanks to the coherent combination of all antenna
signals.  This results in significantly reduced output specifications
of the Power Amplifiers (PAs). The power amplification stage typically
accounts for $>70\%$ of the power consumption of base stations in
wireless broadband macro-cells \cite{Auer2011}. Moreover they
necessitate cooling, causing a $\sim 10\%$ overhead. The
reduced output power in Massive MIMO hence can reduce the total power
by a factor of 3 in an exemplary 100-antenna base station, assuming
that all other contributions remain equal.

The PA mostly operates at a low efficiency as a consequence of a
considerable back-off, required to avoid entering the saturation region.  For
OFDM-based systems such as 3GPP-LTE, the PA typically operates
with a back off of 8--12~dB. Best-in-class solutions need complex
techniques that achieve an efficiency of $\sim 30\%$ \cite{Li2011}.
Entering the saturation region introduces non-linear distortion, which
 comes with two detrimental effects: distortion of the intended
signal within the band of interest, and out-of-band (OOB) emissions
that result in adjacent channel leakage.

We consider a polynomial memoryless model \cite{Horlin2008} for the
non-linear behaviour of the PA. The impact on the signal at RF can be
expressed as:
\begin{equation}
y(t) = \sum_{p} \alpha_p {x}^p_{\text{RF}}(t),
\end{equation}
where $x_{\text{RF}}(t)$ is the input signal to the PA, $y(t)$ is the output
signal, and $\alpha_p$ is the non-linear distortion coefficient of the
PA for the $p$th harmonic component. The third-order harmonic will
have the largest impact both in terms of in-band distortion and
adjacent channel leakage. Furthermore, the amplitude will be limited
to the saturation amplitude $a_{out,sat}$ for input values exceeding
the input saturation amplitude $a_{in,sat}$:
\begin{equation}
\left|y(t)\right| = a_{\text{out,sat}}; \quad  \left|x_{\text{RF}}(t)\right| > a_{\text{in,sat}.}
\end{equation}

The non-linear distortion resulting from the PAs in the many antenna
paths is hence signal dependent. The input signals to the PAs can be
correlated, depending on the specific communication scenario in terms
of users, channel responses, and power (im)balance among the
users. In \cite{Larsson2017PA} we analyzed how the distortion terms
can combine by means of a basic dual-tone modulation scheme. The
following effects can occur:
\begin{enumerate}
\item The distortions may add up coherently in the channel and
	generate considerable out-of-band emissions. This will be the
	case for example in a single-user situation with one strongly
	dominating propagation direction. 

\item In most multi-user scenarios
	the precoder will provide significant different compositions
	of signals to the antenna paths and hence power amplifiers. In
	general, this will randomize the harmonic distortion terms.
\end{enumerate}
 
\begin{figure}[t!]
\centering
\includegraphics[width=0.5\textwidth]{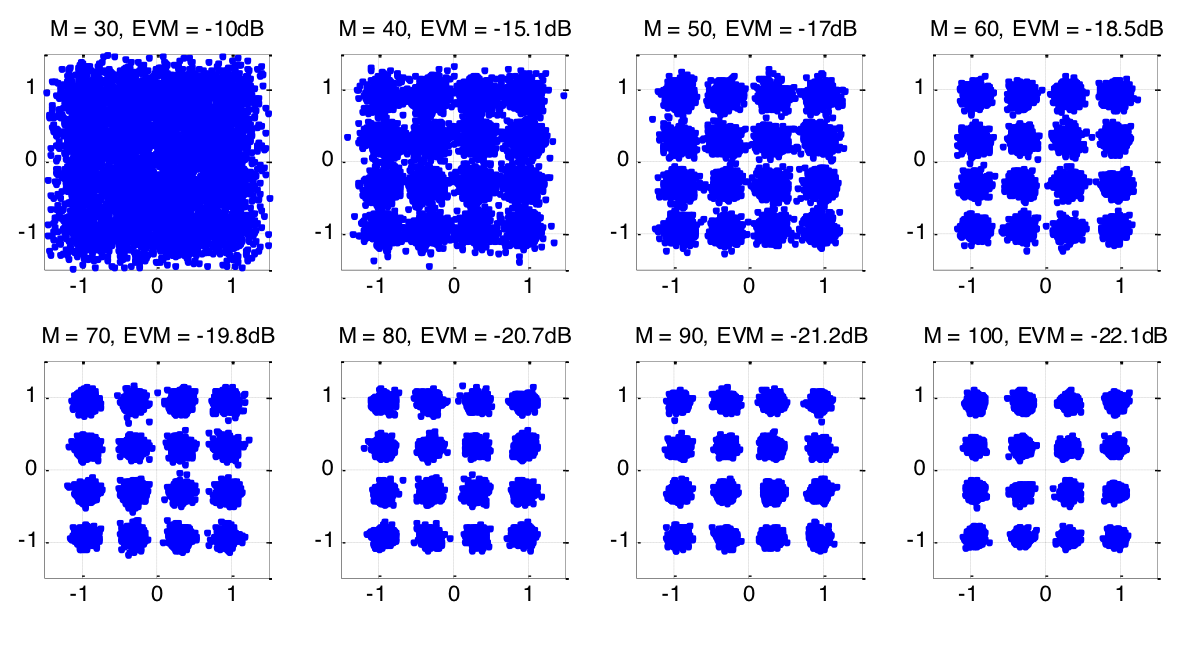}
  \caption{Increasing number of base station antennas improves the EVM with PAs operating in saturation.\label{fig:PAconstellation}}
\end{figure}

The constellation diagrams in Figure~\ref{fig:PAconstellation}
illustrate the impact of increasing the number of antennas at the base
station on the Error Vector Magnitude (EVM),
 for a
case with equal-strength signals for the different users and i.i.d.\
Rayleigh fading channels. The results were simulated based on a cubic
polynomial model for the PA, which operates in saturation ($0$~dB with
respect to the $1$ dB compression point).  With $M=30$ antennas at the
base station, the constellation points are seriously dispersed and an
EVM of $-10$ dB is measured. When increasing the number of antennas,
in steps of 10 in the graph, the clarity of the constellation diagram
greatly improves and for $M=100$ an EVM of $-22$ dB is observed.

In conclusion, the power amplifiers benefit from the large array owing
to the drastically reduced total output power requirement. Moreover in
many typical conditions, Massive MIMO systems will not transmit
predominantly to one user and in one direction. One could then operate
the PAs efficiently in their non-linear region.  Hence, a considerable
further improvement of the power consumption could be achieved.
However, the inconvenient truth is that in general, directive
emissions of OOB radiation can arise under some conditions.  More
detailed mathematical models and results can be found
in \cite{DBLP:journals/corr/abs-1711-02439}.

\subsection{Coarse and Lean Convertors}

The impact of low-resolution data converters on  system performance has been investigated. We give an overview of these theoretical results and discuss them in perspective of actual design constraints and merits of state-of-the-art data converters. These reveal that minimizing the resolution strictly (e.g., below 6 bits) does not result in a significant power reduction in a conventional base station. One should hence question any penalty in system performance and/or additional DSP complexity when considering very low resolution data converters.

A specific type of hardware distortion arises if low-precision A/D
converters are used at the base station. Such converters are highly
desirable owing to their low cost and power consumption. In principle for each bit reduction in resolution, the A/D converter power is halved. Doubling the sampling frequency
 will double the power. This is reflected in the common figure-of-merit (F.o.M.) in terms of energy consumption per conversion step (cs) \cite{Pelgrom17book} used to assess the design merit of A/D converters implementing different architectural principles and resolution/bandwidth specifications:
\begin{equation}
\text{F.o.M.}_{\text{A/D}}= \frac{\text{Power}_\text{A/D}}{2^{\text{ENOB}}\cdot f_{s}}
\label{eq:FOMADC}
\end{equation}
where ENOB is the Effective Number of Bits resolution as measured and $f_{s}$ is the sampling frequency.

The
resulting quantization noise of A/D conversion is fairly easy to model accurately, and
rigorous information-theoretic analyses of its effect are available. In some cases, line-of-sight with a single terminal, the quantization
noise may combine constructively.  However, in frequency-selective,
Rayleigh fading channels with large delay-spreads and multi-user
beamforming, the distortion averages out over the
antennas to a significant extent.  Specifically, with 1-bit quantization, the quantization
noise has a power equal to $(\pi/2-1)P$ where $P$ is the received
signal power \cite{mollen1bit}, and the aggregate effect of the
quantization is approximately a loss in effective SINR of 4~dB. 
The 1-bit A/D converter case is of particular interest as it allows operation without automatic
gain control (AGC), which simplifies hardware complexity. 
With
$N$-bit quantization, $N>1$, corresponding results  can be found
in \cite{mollen2016achievable}, and when $N$ grows eventually the
capacity formulas for the un-quantized
case \cite[Ch.~3]{Marzetta16book} are rediscovered.  Other authors
have derived similar results subsequently \cite{7931630} -- and
earlier, using heuristic
arguments, \cite{fan2015uplink,zhang2016spectral}.  Importantly, these
analyses take into account the fact that both the received pilots and
the payload data will be affected by quantization noise.  
 
The loss in effective SINR due to quantization needs to be considered relatively to the extra power consumption resulting from adding bits resolution in the A/D converters.  
Circuit innovation in data converters has brought great improvements in power efficiency. State-of-the-art designs for A/D converter cores achieve figures-of-merit following \ref{eq:FOMADC} in the order of $10$ fJ/cs \cite{vanderplas2008,Choo2016}. 
A 6-bit ADC with a speed of several 100 Mbit/s  consumes $<1$~mW.  

Massive MIMO systems operating with low-resolution Digital-to-Analog (D/A) converters at the base station in the downlink transmission have also been studied. There is some evidence that they are sufficient to attain a good performance in terms of achievable link rate \cite{7887699,Jacobsson2017rate}. Also, while
these analyses are independent of the actual modulation and coding
used in the system, numerical end-to-end link simulations have
independently arrived at essentially the same
conclusion that the degradation of BER performance due to low-precision ($<6~bits$) D/A converters is negligible \cite{Desset2015}. It is however a misconception that the number of bits resolution affects the D/A converter power consumption in a similar way as it does for A/D converters. The constraint on OOB emission in combination with the swing to be delivered to the analog output signal are the dominant factors in the power and complexity of a D/A converter \cite{Pelgrom17book}. A relevant standard figure-of-merit (F.o.M.)  for current-steering D/A converters is given by
 \begin{equation}
 \text{F.o.M.}_{\text{D/A}}= \frac{V_{\text{pp}}\cdot f_{\text{out}}\cdot 10^{\text{SFDR}/20}}{\text{Power}_{\text{D/A}}}
 \label{eq:FOMDAC}
 \end{equation}
where SFDR is the spurious free dynamic range, being the distance between the signal and the largest single unwanted component -- the spurious signal,
and $V_{pp}$ is the peak-to-peak signal swing which accounts for the
power (and design problems) needed for generating the analog signal in a digital-to-
analog converter. D/A converters with a resolution $<10$~bits are conveniently implemented by current injection or resistive architectures whose power consumption is typically not directly impacted by their resolution.
In contrast,  the complexity of the reconstruction filter in the D/A converter is mostly determined by the SFDR specification, which will eventually determine the out-of-band (OOB) harmonic distortion. Digital predistortion and analog filtering to reduce OOB emissions have been proposed for coarsely quantized precoding in Massive MIMO \cite{Jacobsson2017}. The extra processing complexity in deeply scaled technology will be very reasonable, yet a degradation of the in-band signal-to-interference-noise-and-distortion ratio (SINDR) on the link is introduced. This presents the same trade-off between in-band transmission versus out-of-band rejection  encountered in D/A converter design. 

The trend in broadband wireless systems to increase spectral efficiency through a combination of higher order modulation constellations and conventional multi-layer MIMO has raised the resolution requirement for data converters $>\sim 12$~bits. Massive MIMO can operate without noticeable implementation loss with only $4-6$-bit A/D and D/A converter resolution. This reduces the power consumption of an individual A/D converter specifically with a factor $>\sim 50$, which more than compensates for the fact that $10-30$ times more converters are needed. 
It is however neither necessary nor overall beneficial to reduce the resolution of A/D and D/A converters below 6~bits:
\begin{itemize}
	\item On uplink, reducing the A/D resolution further will save less than $100$~mW in a 100~antenna basestation. 
	
	\item On downlink, a potential implementation loss of $0.5$~dB  or more due to a D/A converters with a lower resolution may require $10\%$ more power in the PA stage. More importantly, the constraints on OOB emission will not be met. Dedicated processing will hence be needed to avoid or filter out unacceptable leakage in adjacent bands.  
\end{itemize}   

\subsection{Reciprocity Calibration   in RF Front-Ends}

Channel estimates are obtained
from uplink pilots; see   Section~\ref{sec:concept}. In
 practice, the response observed by the digital
baseband processing for each user includes both the propagation
channel and the transceiver transfer functions. The full
responses for uplink and downlink can be expressed as:
\begin{equation}
\begin{array}{l}
\bg_{k,UL}=\bR_{B} \tilde{\bg_{k}} t_{k} \\
\bg_{k,DL}^T=r_{k} \tilde{\bg_{k}}^T \bT_{B},
\end{array}
\end{equation}
where $\bR_B$ and $\bT_B$ are complex diagonal matrices containing the
base station receiver and transmitter responses, and $t_k$ and $r_k$
are the responses of the transmitter and receiver of user terminal
$k$.  While the responses of the propagation channel $\tilde\bg_{k}$
are reciprocal, the responses of the front-ends will typically cause
non-reciprocity in the full response. In the precoded Massive MIMO
downlink reception the following holds:
\begin{equation}
\begin{array}{l}
 \bR_{B} \neq \bT_{B}\\
 r_{k} \neq t_{k}\\
\Rightarrow \bg_{k,DL} \neq \bg_{k,UL}.
\end{array}
\end{equation} 
When the corresponding estimates $\hat\bg_{k}$ of $\bg_{UL}$ are used
to calculate the precoding coefficients, they will introduce
Multi-User Interference (MUI) and potentially an SNR loss, depending on
the precoding vectors $\ba_k$.  We include the derivation for the
zero-forcing precoder, and refer to \cite{MAMMOETD2_4} for a
comprehensive treatment. Under the assumption of negligible channel
estimation errors and considering normalized responses to simplify
notation,\footnote{Power control does not  impact
reciprocity, and it will show up as a scalar multiplication on the
individual terminal signals.} the received signals at the terminals
$\by=[y_1,\ldots,y_K]^T$ are given by
\begin{equation}
\by = \bG_{DL}^T \bG_{UL}^*(\bG_{UL}^T\bG_{UL}^*)^{-1} \bx + \bw
\end{equation}
where $\bx$ and $\bw$ are the $K$-vectors of transmitted symbols and received
noise samples, respectively. Writing out the front-end responses gives
the following expression:
\begin{equation}\label{eq:recip}
\by = (\bR_{U} \tilde\bG^T \bT_{B}) (\bR_{B}^{*}\tilde\bG^*\bT_{U}^*)(\bG_{UL}^T\bG_{UL}^*)^{-1} \bx + \bw,
\end{equation}
where $\bR_{U}$ and $\bT_{U}$ are diagonal matrices containing the
transmitter and receiver responses of terminals $t_{k}$ and
$r_{k}$. Equation (\ref{eq:recip}) shows that in general the combined
precoding, channel, and transceiver responses will not result in a
diagonal matrix. As a result, MUI will occur.  Structurally it is the
multiplication of the base station's front-end responses
$\bT_{B}\bR_{B}^*$ that is responsible for the MUI. The terminal
responses appear as scalar multiplications on the received symbols and
will be contained in the equalization processing in the terminal. A
suitable calibration procedure operating locally at the base station
can restore the reciprocity.  Calibration data needs to be obtained
through measurements of the transceiver front-end responses, for which
several approaches have been proposed and validated:
\begin{itemize}
\item Utilization of
an auxiliary front-end, which sequentially measures the RF transceiver
front-ends. The method works well in conventional MU-MIMO
systems \cite{Bourdoux2003}. However, it does not scale well to large
numbers of antennas.

\item Exploitation of the coupling, essentially radio propagation, between antennas in
the array to derive the relative differences among the transceiver
responses. This solution has been implemented in real-life testbeds
and performs well \cite{Vieira2017}.
\end{itemize} 

Analysis has shown that non-reciprocity
requirements are not as severe for Massive MIMO as in conventional systems \cite{MAMMOETD2_4} and depend on the system
load and precoding algorithms.  The RF transceiver responses may vary
in time  mainly due to temperature differences. The calibration
procedure hence needs to be repeated on a regular basis. In typical
conditions the required updating frequency is in the order of
hours. It thus introduces only very limited overhead.

%% file: precode_decode.tex
The central detector and precoder perform the crucial operations to
achieve   spatial multiplexing. This section zooms in on the hardware
implementation of the precoding and detection algorithms.

\begin{table*}[t]
\renewcommand{\arraystretch}{1.5}
\caption{Computational complexity (number of real multiplications) of different detection techniques.}
\label{tb:comp} \centering
\begin{tabular}{|c|c|c|}
\hline
Algorithms & Per channel realization & Per channel use \\
\hline
Neumann Series & $2MK(K+1)+8K^2+4(L-1)K^3$ & $4K^2+4KM$\\
\hline
Cholesky Decomp. & $2MK(K+1)+4K(K+1)(K+2)/6$  & $4K^2+4K+4KM$ \\
\hline
Modified QR Decomp. & $2MK(K+1)+4K^3/3+3K^2/2-31K/3$  & $6K^2-2K+4KM$\\
\hline
Coordinate Descent & - & $4M(L-1)+4KML$ \\
\hline

\end{tabular}

\end{table*}

\subsection{Implementation Challenges and Design Considerations}

Linear processing provides good precoding and detection performance
under favorable propagation conditions. However, linear processing in
Massive MIMO does not necessarily result in low computational
complexity given that the operations need to be performed on large
matrices.  For instance, the complexity of computing $(\bG^H\bG)^{-1}$
for an $M\times K$ matrix $\bG$ is
\begin{equation}
MK^2+K^3.
\end{equation}
This number is in the order of $10^4$ for an $M=128$, $K=16$
system. In TDD massive MIMO systems, processing latency is a crucial
design consideration, especially for high-mobility scenarios. The
analysis in \cite{Steffen2016SIPS} shows that the time budget for
operating the precoding is 150 $\mu$s to support a moderate mobility
of $70$ km/h. The high computational complexity and processing speed
need to be handled with reasonable hardware cost and power
consumption. These implementation challenges necessitate meticulously
optimized solutions following a systematic algorithm-hardware
co-design methodology.

A central property of Massive MIMO is that the column vectors of the
channel matrix are asymptotically orthogonal under favorable
propagation conditions. As a result, the Gram matrix, $\bZ=\bG^H\bG$,
becomes diagonally dominant, i.e.,
\begin{equation}
|z_{i,i}|\gg |z_{i,j}|, \text{for } i\neq j \text{ and } M\gg K,
\end{equation}
and for i.i.d.\ channels,
\begin{equation}
\frac{1}{M}\bZ \rightarrow \bI, \text{for } M\rightarrow\infty \text{ and for fixed } K.
\end{equation}
The extent of the diagonal dominance varies with the
characteristics of the antenna array, the propagation environment, and
the number of users served. Exploiting this dominance, approximate matrix
inversion can be performed to reduce the computational complexity.
Matrix inversion approaches can be categorized into three types:
explicit computation, implicit computation, and hybrid methods. We
next assess the complexity and suitability of these methods.

\subsection{Explicit Matrix Inversion}

Explicit matrix inversion can be performed using approaches such as
Gauss-elimination, Neumann series expansion
\cite{ZhuICC2015}, and truncated polynomial expansion
\cite{Mueller2016EURASIP}. Recently, the Neumann series approximation
has been identified as one of the most hardware-friendly algorithms for
Massive MIMO systems \cite{Hemanth2014ISCAS,Wu2014}.  If a $K\times K$
matrix $\bZ$ satisfies
\begin{equation}
\lim_{n \to \infty} ( \bI - \bX^{-1}\bZ)^n \simeq \mathbf{0}_K,
\end{equation}
its inverse can be approximated by a Neumann series with $L$ terms as:
\begin{equation}
\bZ^{-1} \approx \sum_{n=0}^{L} \left( \bI - \bX^{-1} \bZ \right)^n \bX^{-1},
\label{eq:neuman_inv}
\end{equation}
where $\bX$ is a pre-conditioning matrix. The number of terms, $L$, can
be used as a tuning parameter to trade off between complexity and accuracy. It
is shown in \cite{Wu2014} that using the main diagonal of the Gram
matrix,
\begin{equation}
\bZ_d=\text{diag}[\bZ_{1,1},\cdots,\bZ_{K,K}],
\end{equation}
as the pre-conditioning matrix, the Neumann series approximation can
provide close-to-exact-inversion performance with $L=3$ when $K\ll
M$. However, a significant performance loss is demonstrated when $M/K
< 8$. To improve the accuracy, the following weighted Neumann series
approximation was introduced in \cite{Lee2016tvt,Nagy2017WCL}:
\begin{equation}
\bZ^{-1} \approx \sum_{n=0}^{L} \alpha_n \left( \bI - \bX^{-1} \bZ \right)^n \bX^{-1}.
\label{eq:neuman_inv_modified}
\end{equation}
In \cite{Lee2016tvt}, the coefficients $\alpha_n$ are selected by solving the equation
\begin{equation}
\sum_{n=0}^{\infty} \bB^n \approx \sum_{n=0}^{L} \alpha_n \bB^n,
\label{eq:neuman_inv_coef}
\end{equation}
where
\begin{equation}
\bB=-\bZ_d^{-1/2}(\bZ-\bZ_d)\bZ_d^{-1/2}.
\end{equation}
At the price of extra computational complexity, the method in
(\ref{eq:neuman_inv_modified}) improves the performance significantly,
especially in cases with a high user load.

\subsection{Implicit Matrix Inversion}

Implicit matrix inversion uses linear-solvers such as
conjugate-gradient \cite{Yin2014}, coordinate-descent \cite{Wu2016},
and Gauss-Seidel \cite{Gao2015ICC} to perform linear precoding and
detection, without explicitly calculating the Gram matrix inverse. In
\cite{Wu2016}, the coordinate-descent method is adopted to realize an
MMSE   detector. The regularized squared Euclidean distance,
\begin{equation}
f(\bx)= \|\by-\bG\bx\|_2^2+N_0\|\bx\|_2^2,
\label{eq:mse}
\end{equation}
is minimized sequentially for each variable in $\bx$ in a round-robin
fashion. In (\ref{eq:mse}), $N_0$ is the variance of each complex entry in the noise
vector $\bw$. In each iteration, the solution for the $i$th element in $\bx$ is
\begin{equation}
\hat{x}_i= \frac{1}{\|\bg_i\|_2^2+N_0}\bg_i^H \pp{\by-\sum_{j\neq i}\bg_jx_j}.
\label{eq:code}
\end{equation}
This procedure is then repeated for $L$ iterations.

\subsection{Hybrid Method}

Matrix decomposition algorithms factorize the Gram matrix into
intermediate matrices, which are generally triangular. Forward or
backward substitution is then performed to accomplish the
corresponding precoding and detection operation.  The solution in
\cite{Prabhu2017} utilizes QR-decomposition. The Gram matrix $\bZ$ is
decomposed as
\begin{equation}
\bZ=\bQ\bR,
\end{equation}
where $\bQ$ is unitary and $\bR$ is upper triangular. The linear
equation $\hat{\bs}=\bZ^{-1}\bs$ is then rewritten as
\begin{equation}
\bR\hat{\bs}= \bQ^H\bs,
\end{equation}
which can be solved using backward substitution. This method avoids
the explicit computation of matrix inverses, relaxing (to some extent)
the requirements on data representation accuracy.  By exploiting the
diagonally dominant property of the Gram matrix, modified
QR-decomposition can be performed \cite{Prabhu2017}. For instance, the
original solutions
\begin{equation}
\begin{array}{l}
c = a/r  \\
s =b^*/r\\
r=\sqrt{|a|^2+|b|^2}
\end{array}
\label{eq:qr_givens}
\end{equation}
to the Givens rotation operation
\begin{equation}
\left[
\begin{array}{cc}
  c & s \\
  -s^{*} & c
\end{array}
\right]
\left[
\begin{array}{c}
a\\
b
\end{array}
\right]=
\left[
\begin{array}{c}
r\\
0
\end{array}
\right],
\label{eq:qr_givens_val}
\end{equation}
are approximated by
\begin{equation}
\begin{array}{l}
c=c_{\textrm{const}} \\
s=b^*/a.
\end{array}
\label{eq:qr_givens_approx}
\end{equation}
Equation (\ref{eq:qr_givens_approx}) makes use of the fact that
$|a|\gg |b|$ and results in 50\% complexity savings by introducing the
constant $c_{\textrm{const}}$.

Cholesky-decomposition ($\bZ=\bL\bL^*$) has also been studied for
Massive MIMO precoding and detection implementation
\cite{Alshamary2016ISIT,Rakesh2017ISCAS}. It has lower computational
complexity than the Neumann series expansion method (with $L\geq 4$)
\cite{Wu2014} and provides accurate processing independent of $M$ and
$K$. More importantly, the Cholesky decomposition imposes lower
memory requirements, since only the lower triangular
matrix $\bL$ needs to be stored.

\subsection{Complexity versus Accuracy Trade-Off}

\begin{figure}[t]
\centering
\includegraphics[width=0.38\textwidth]{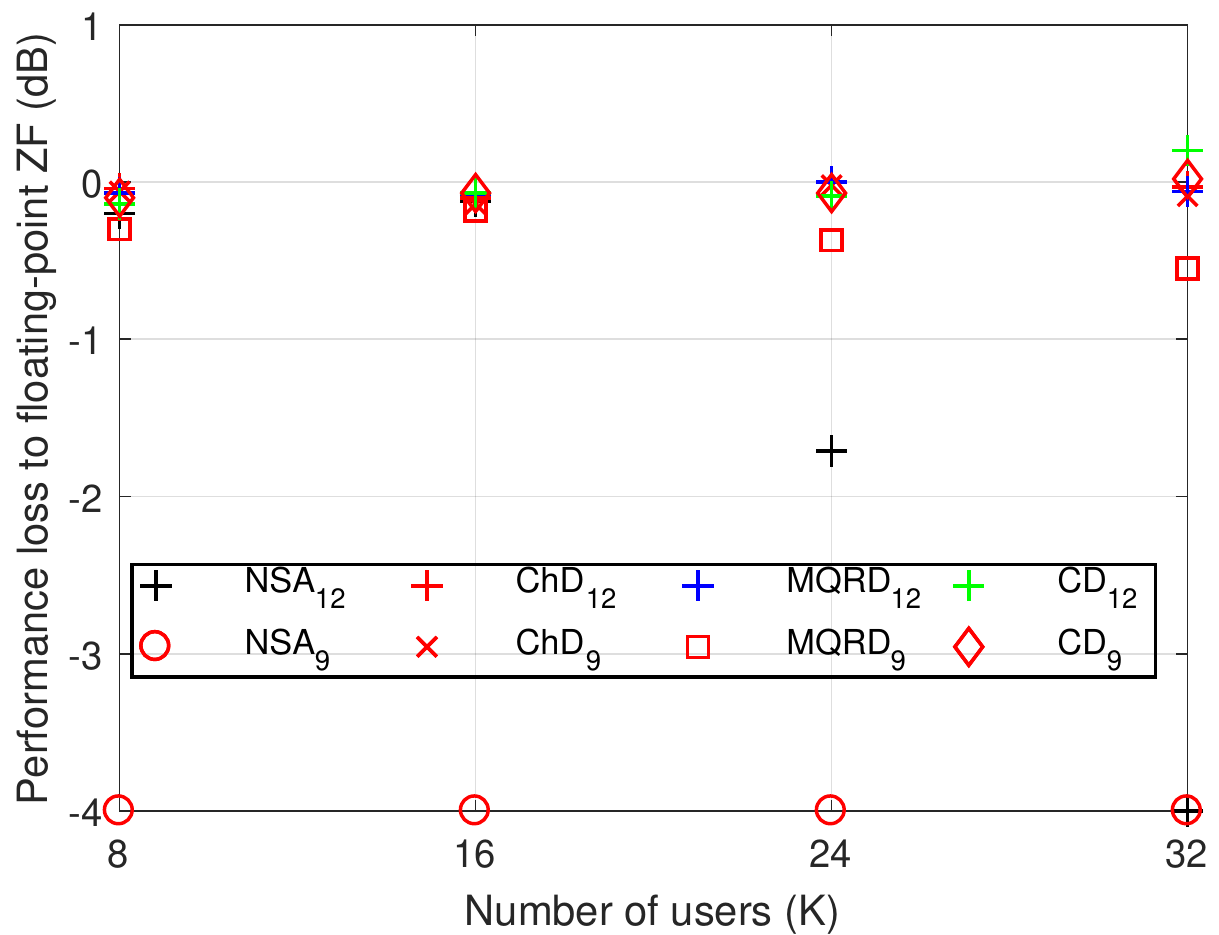}
\caption{Simulated performance of different detection methods. The
  subscripts in the legend indicate the fixed-point resolution of the
  fractional part. Markers at $-4$ dB performance loss mean that the corresponding
  detection scheme has a performance loss greater than $4$ dB or shows
  an error floor before reaching a BER of $10^{-4}$.}
\label{fig:performK}
\end{figure}

To select appropriate processing algorithms for Massive MIMO is
non-trivial, and an analysis of the trade-off between computational
complexity and processing performance is necessary. Reference
\cite{Mirsad2014} presents such an analysis for different MMSE
detection techniques.

To evaluate the processing accuracy, we simulate the performance of
different detection techniques including Neumann series approximation
(NSA), Cholesky decomposition (ChD), modified QRD (MQRD), and
coordinate descent (CD). The effects of fixed-point arithmetics is
also taken into consideration to examine the required data
precision. In the simulations, $M=128$, $K$ sweeps from 8 to 32, and
an i.i.d.\ block Rayleigh fading channel with perfect channel
estimation and synchronization  was considered.
 A rate-$1/2$ convolutional code with generator polynomial
[171, 133] and a constraint length of 7 was used. Figure~\ref{fig:performK} shows the
performance at $10^{-4}$ BER relative to floating-point
ZF detection. The number of iterations $L$ for the NSA and CD was set
to 3. Implicit and hybrid methods are more robust to lower
resolutions, while NSA requires a larger number of bits to calculate
the matrix inverse explicitly.  When $M/K$ is small the Gram matrix
becomes less diagonally dominant and approximate matrix inversion
methods suffer from a larger performance loss. 
CD offers better interference cancellation when the user load is relatively high.

\begin{figure}[t]
\centering
\subfigure{
\includegraphics[width=0.38\textwidth]{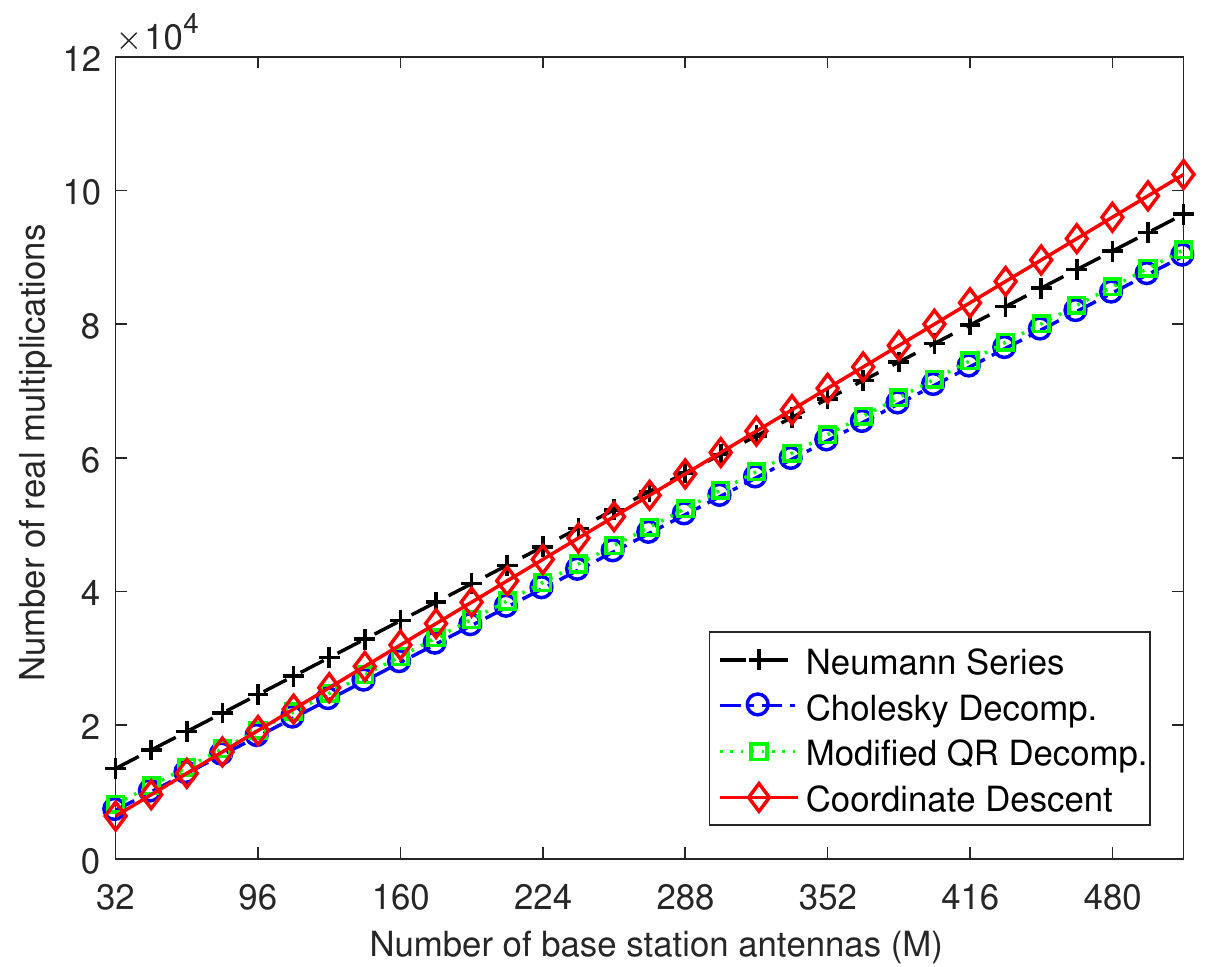}
}
\subfigure{
\includegraphics[width=0.38\textwidth]{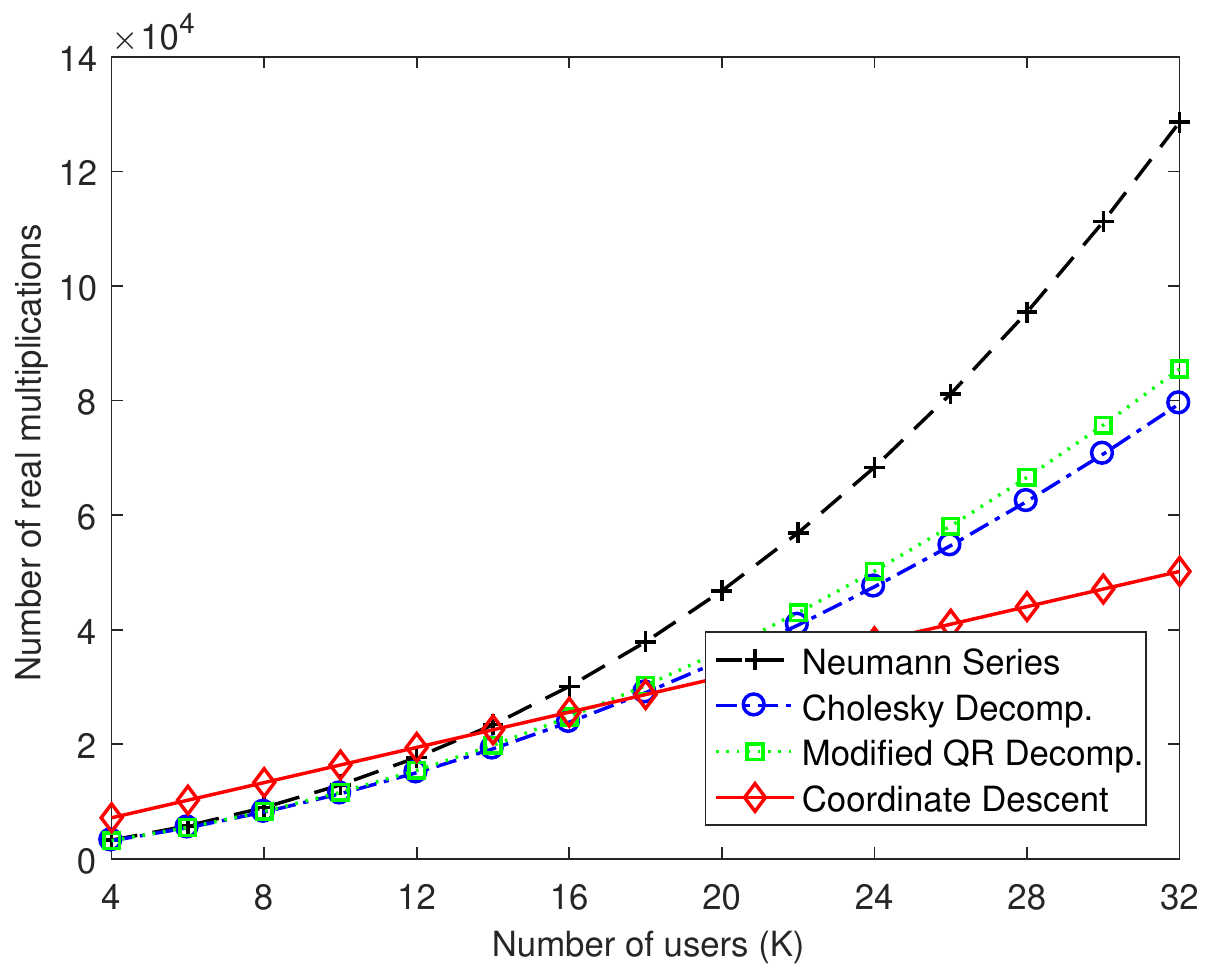}
}
\subfigure{
\includegraphics[width=0.38\textwidth]{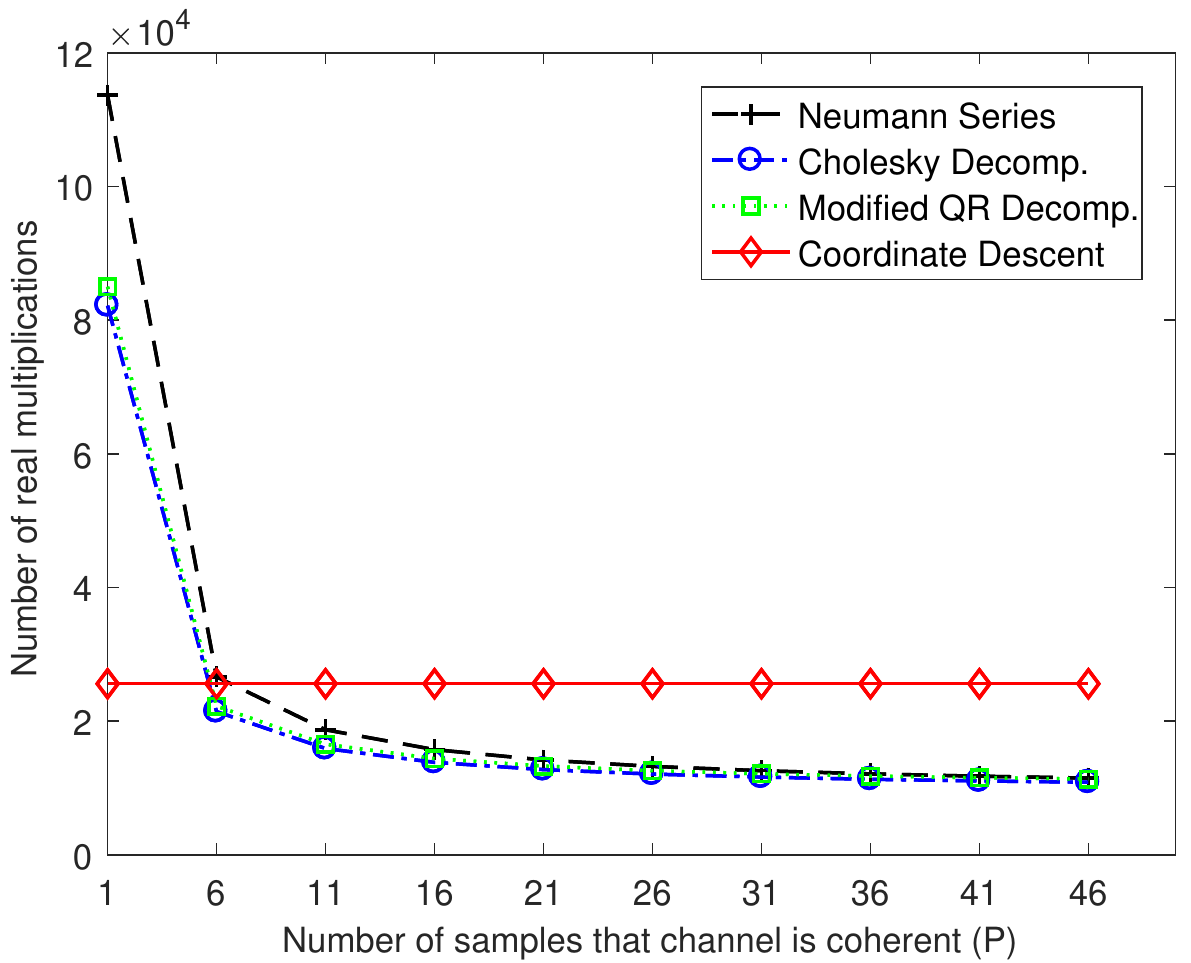}
}
\caption{Computational complexity (per instance of the detection
  problem) of different implementations of ZF detection, for different
  numbers of base station antennas, numbers of users, and channel
  coherence duration.}
\label{fig:comp}
\end{figure}

  Table \ref{tb:comp}  lists the corresponding computational
complexity in terms of   number of real multiplications. The
computation is divided into two parts depending on how frequently it
needs to be executed, i.e., per channel realization and per channel
use (instance of the detection problem). The Gram matrix calculation,
matrix decomposition, and matrix inversion are performed when the
channel changes, while matched-filtering and backward/forward
substitution are performed for each received vector. Thereby, the
computational complexity depends on the channel dynamics, i.e., the
number of samples ($P$) during which the channel is constant.
Figure~\ref{fig:comp} depicts the results. Different system setups and
channel conditions are analyzed. While changing $M$, $K$, and $P$ in
the three sub-figures, the other two are fixed to $M=128$, $K=16$, and
$P=5$, respectively. Several observations can be made. The detection
complexity grows linearly with $M$, enabling
large savings in   transmit power by deploying
large numbers of antennas, with a mild increase in the processing
power. Moreover, the processing complexity (for explicit and hybrid
matrix inversion algorithms) can be dramatically reduced in static
environments, in which case the channel matrix-dependent operations
are performed very rarely.

In addition to the processing accuracy and computational complexity,
parallelism is an important aspect to be considered, and it
highly impacts the processing latency. Iterative algorithms such as
Neumann series approximation and coordinate descent can suffer from a
long processing latency for MUI-dominant channels. On the other hand,
matrix decomposition can be performed in a more parallel fashion and
was thus selected for the first Massive MIMO precoder-detector chip
introduced in the next section. Moreover, the intermediate results
$\bZ^{-1}$, $\bL$, and $\bQ\bR$ can be shared between the uplink and
downlink processing, further simplifying the hardware.

\subsection{128$\times$8 Massive MIMO Precoder-Detector Chip Achieving 300 Mb/s at 60 pJ/b}

\begin{figure*}[t]
\centering
\includegraphics[width=0.8\textwidth]{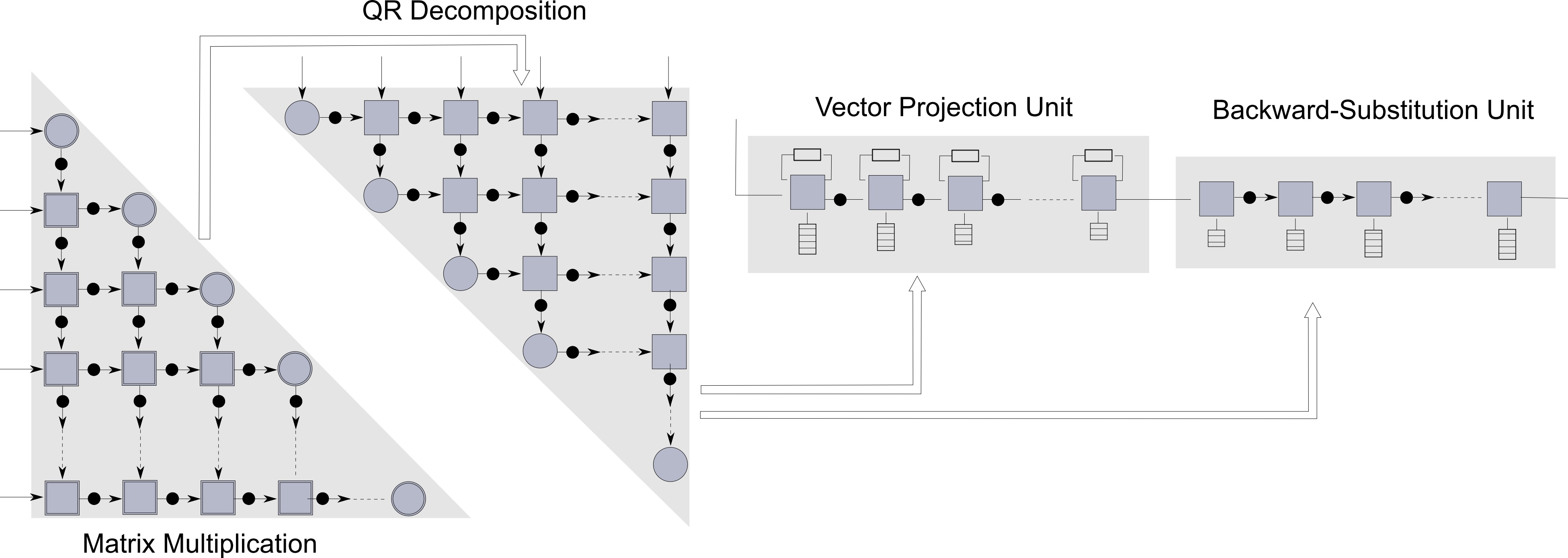}
\caption{Simple, configurable, and scalable architecture for QRD-based massive MIMO precoder (From \cite{Prabhu2017}).}
\label{fig:arc}
\end{figure*}

Integrated hardware implementations will ultimately define both the
performance and power consumption of Massive MIMO systems. Hence,
algorithms should be selected such that the corresponding operations
can be mapped into simple, configurable, and scalable hardware
architectures to enable high throughput, low latency, and flexible
implementation. The reconfigurability and scalability are essential to
enable efficient operation in a wide range of conditions. In this
section we present a design \cite{Prabhu2017} demonstrating such an
algorithm and hardware architecture co-design, where the
QR-decomposition based ZF precoding is mapped onto a systolic array
architecture; see Figure~\ref{fig:arc}. The systolic array consists of
a homogeneous network of elementary processing nodes, where each node
performs the same pre-defined tasks. Due to the homogeneity, the
architecture is scalable to support different $M$ and $K$.
The data flow in a systolic array is straightforward and parallel, leading to a simple and
high-speed hardware implementation.

\begin{figure}[t!]
    \centering
    \begin{subfigure}[]
        \centering
        \includegraphics[width=0.20\textwidth]{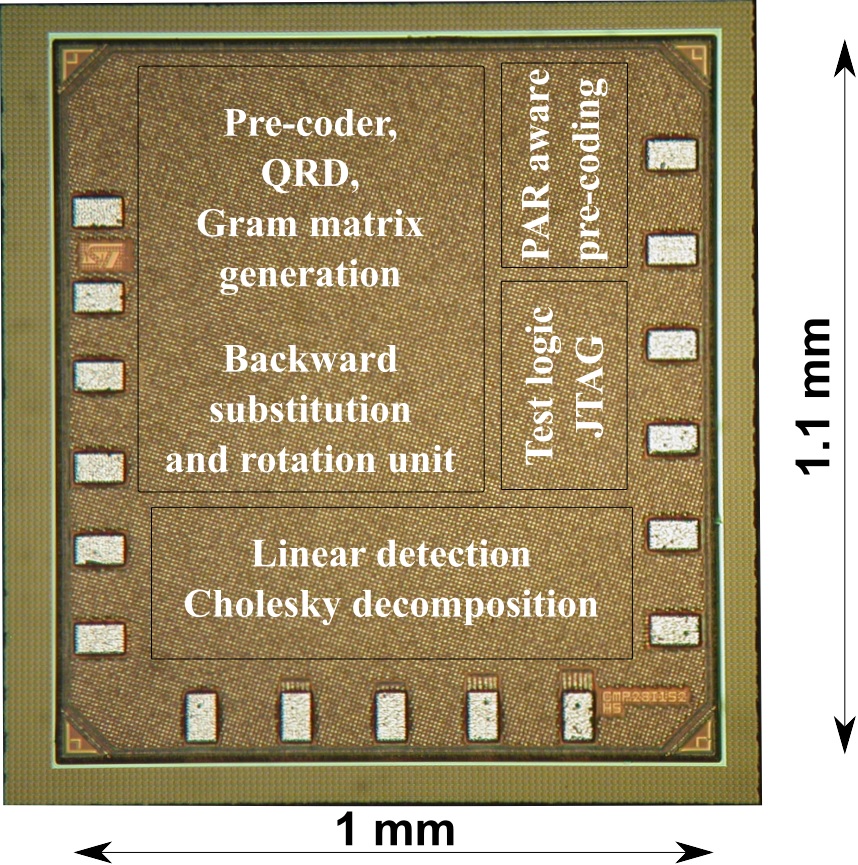}
    \end{subfigure}
    \begin{subfigure}[]
        \centering
        \includegraphics[width=0.20\textwidth]{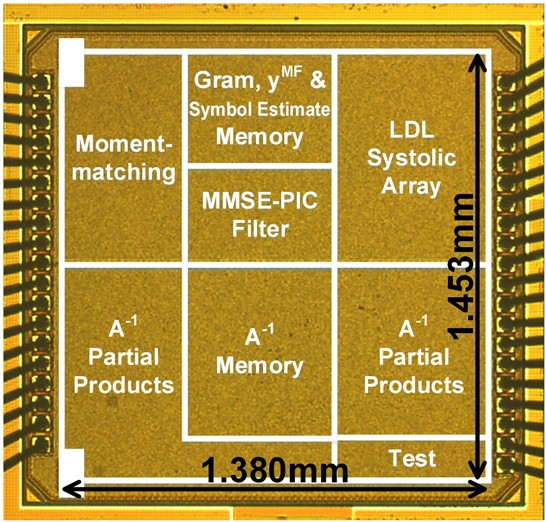}
    \end{subfigure}
    \caption{Microphotographs of massive MIMO precoder and detector chips: (a) From \cite{Prabhu2017} (b) From \cite{Tang2018}.}
\label{fig:chip}
\end{figure}

The QR-decomposition based precoder, together with a
Cholesky decomposition based detector, was fabricated using $28$ nm
FD-SOI (Fully Depleted Silicon On Insulator)
technology. Figure~\ref{fig:chip}(a) shows a photograph of the chip. It
occupies only a $1.1$ mm$^2$ silicon area and consumes $\sim~50$~mW 
power for precoding and detection for 
a 128$\times$8 Massive MIMO system with a $300$ Mb/s throughput. The
fabricated chip and the measurement results prove that the Massive
MIMO concept works in practice and that system-algorithm-hardware
co-optimization enables record energy-efficient signal processing. The
cross-level design approach also applies advanced circuits techniques
leveraging on the flexible FD-SOI body bias feature
\cite{Nicolas2012}. Using forward body bias or reverse body bias
allows systems to dynamically adjust    processing speed and power
consumption of the chip towards the most efficient operating point.

The algorithm-hardware co-design method is further exploited in \cite{Tang2018} to map an iterative expectation-propagation detection (EPD) onto a condensed systolic array for higher hardware resource utilization. This detector chip (Figure~\ref{fig:chip}) is   fabricated using $28$ nm FD-SOI technology and provides 1.8~Gb/s throughput with $127$~mW power consumption. It offers 3 dB processing gain comparing to \cite{Prabhu2017}, equivalent to a 2$\times$ boost in link margin that can be utilized to lower the TX power and relax the front-end requirements.

%% file: PAntP.tex
An obvious concern is how the large number of antennas
and the associated signal processing will affect the cost and energy
consumption of the base station. The individual antenna signals may
have low precision, but regardless of that, the coherent combination
yields excellent SNR eventually.  We demonstrate below that the
resolution of digital signals and operators, such as filtering
coefficients, can be scaled back sharply.  Furthermore, we advocate
processing of the per-antenna functionality without the conventional circuit design margins
that are used  to cope with uncertainties in the semiconductor technology. 
This approach has been called ``at the semiconductor's
edge'' to indicate an operation point where the performance-energy benefit of the technology is maximally exploited at the expense of reliability \cite{Huang2017}. 
Specifically, voltage over-scaling offers significant energy reductions in
deeply scaled CMOS, up to more than 50\%, at the risk of occasional
processing errors. Massive MIMO systems can be designed to meet
required performance levels when operating with error-prone digital
signal processing circuits. Circuits remain functional even for
the worst-case scenario in which the DSP circuitry in some antenna
paths fails completely, for example by a broken power supply. We will
call the situation where the signal in an antenna branch is fully lost
``antenna outage''.  

\subsection{Per-Antenna Functions: Coarse Processing Provides Excellent Performance}
\label{ref:course_PAP}

\begin{figure}[t!]\centering
\includegraphics[width=1\linewidth]{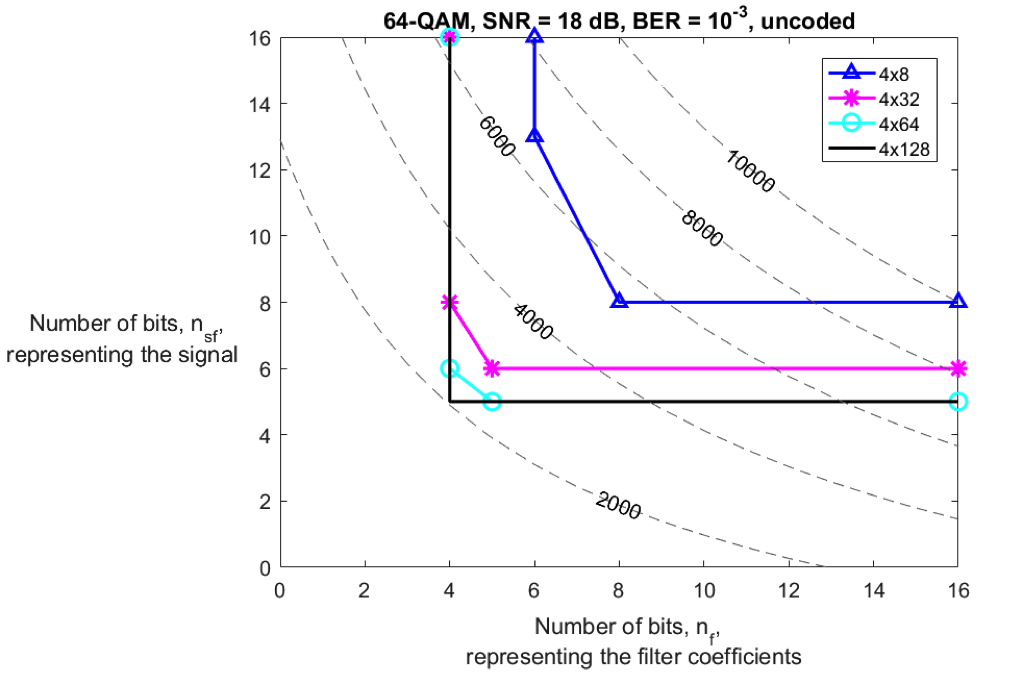}
\caption{Representation of the relative circuit  complexity (area)  as function of the signal
and filter coefficient word lengths. The markers show possible
 operating points with a BER of $10^{-3}$. The dashed lines with
 numbers show operating points with equal complexity. The graphs
 demonstrate that low-resolution processing is feasible with large
 antenna arrays. From \cite{Gunnar2017}.}\label{fig:FilterQuant}
\end{figure}

Massive MIMO can operate well with low-resolution signals. A profiling
of the per-antenna functionality in terms of generic operations per
second shows that for an LTE-like setup, about $80\%$ of the
complexity is in the filtering and the remaining $20\%$ is in the
(I)FFT operation. The filtering functionality is the most demanding
because of the need to over-sample and hence process at high speed.
Significant savings in complexity are therefore possible by minimizing
the resolution of this processing.  An exploration of the word lengths
of the data signals, $n_sf$, and of the filtering coefficients, $n_f$,
is reported on in \cite{Gunnar2017}. The circuit area complexity
$C_\text{Filt}$ of the $T$-tap FIR filtering of I- and Q-signals as a
function of the word lengths is calculated using basic formulas for
the complexity of adders and multipliers, which are dependent on the
word-lengths $n$ and $m$ of the operands as follows:
\begin{equation}
\label{eq:filt_compl}
\begin{array}{l}
C_\text{add} = n\cdot \log_2 n \\
C_\text{mult} = n\cdot m \\
C_\text{Filt} = 2\cdot T\cdot (m+n)\cdot \log_2(m+n)+2\cdot T\cdot m\cdot n.
\end{array}
\end{equation}
If a smaller number of bits is used to represent the signals and the
filter coefficients, the hardware complexity as given in
(\ref{eq:filt_compl}) is reduced. However, decreasing the word length
will increase the quantization noise. For a desired transmission
quality the just-sufficient precision can be determined. Considering
that the quantization noise will be independent among the antennas,
its combined impact will be smaller for larger numbers of
antennas. This effect is illustrated in Figure~\ref{fig:FilterQuant}
for the rather demanding 64-QAM case, and an uncoded Bit-Error Rate
(BER) of $10^{-3}$. The curves were generated based on individual BER vs. SNR simulations for different coefficient and signal resolutions, from which the equal performance points were extracted. Dotted lines show equal-complexity (in terms of
area) solutions.  For a 128$\times$4 Massive MIMO system, 4 and 5 bits
are sufficient for the signals and the coefficients, respectively, for
the targeted performance. This brings a 62\% complexity reduction for
the filters compared to the 8$\times$4 case. The outer right points on the curves are clearly always suboptimal and demonstrate that high-precision filter coefficients do not improve performance, while they can cause a significant complexity penalty. A similar observation holds for the upper left points.
For higher system loads, more bits are needed. At the system level one
could trade-off system load for constellation order to satisfy
throughput requirements.

This analysis provides evidence that low-complexity, coarse processing in the digital filters of the individual antenna signals can offer the required performance in Massive MIMO. In the downlink the signals will next be passed to the D/A converters. The latter could be low resolution as well. The more demanding design challenge for D/A converters however is to meet out-of-band emission specifications, as introduced in Section~\ref{sec:RF}.

The (I)FFT operations required in Massive
MIMO systems with multicarrier modulation can also be designed for
Massive MIMO operation specifically and benefit from the complexity
reduction brought by the coarse quantization. A thorough optimization
is however quite complex and should consider varying quantization at
the different butterfly stages.

\subsection{Processing at the Semiconductor's Edge}
\label{sec:edge}

Applications have benefited over the last decades from Moore's law,
providing ever higher performance at lower power consumption.
Integrated Circuits (ICs) have been able to operate at lower dynamic
power thanks to the scaling of the supply voltage $V_{dd}$. For
digital circuits, the average dynamic power consumption is
\begin{align}
  P_\text{dyn, av} = (\alpha C)\cdot V_\text{dd}^2 \cdot f_s,
\end{align}
where $\alpha C$ is the effective switching capacitance of the module
and $f_s$ is the switching frequency. Clearly, $P_\text{dyn, av}$ scales
quadratically with the supply voltage $V_\text{dd}$.

\begin{figure}[t!]\centering
\includegraphics[width=0.9\linewidth]{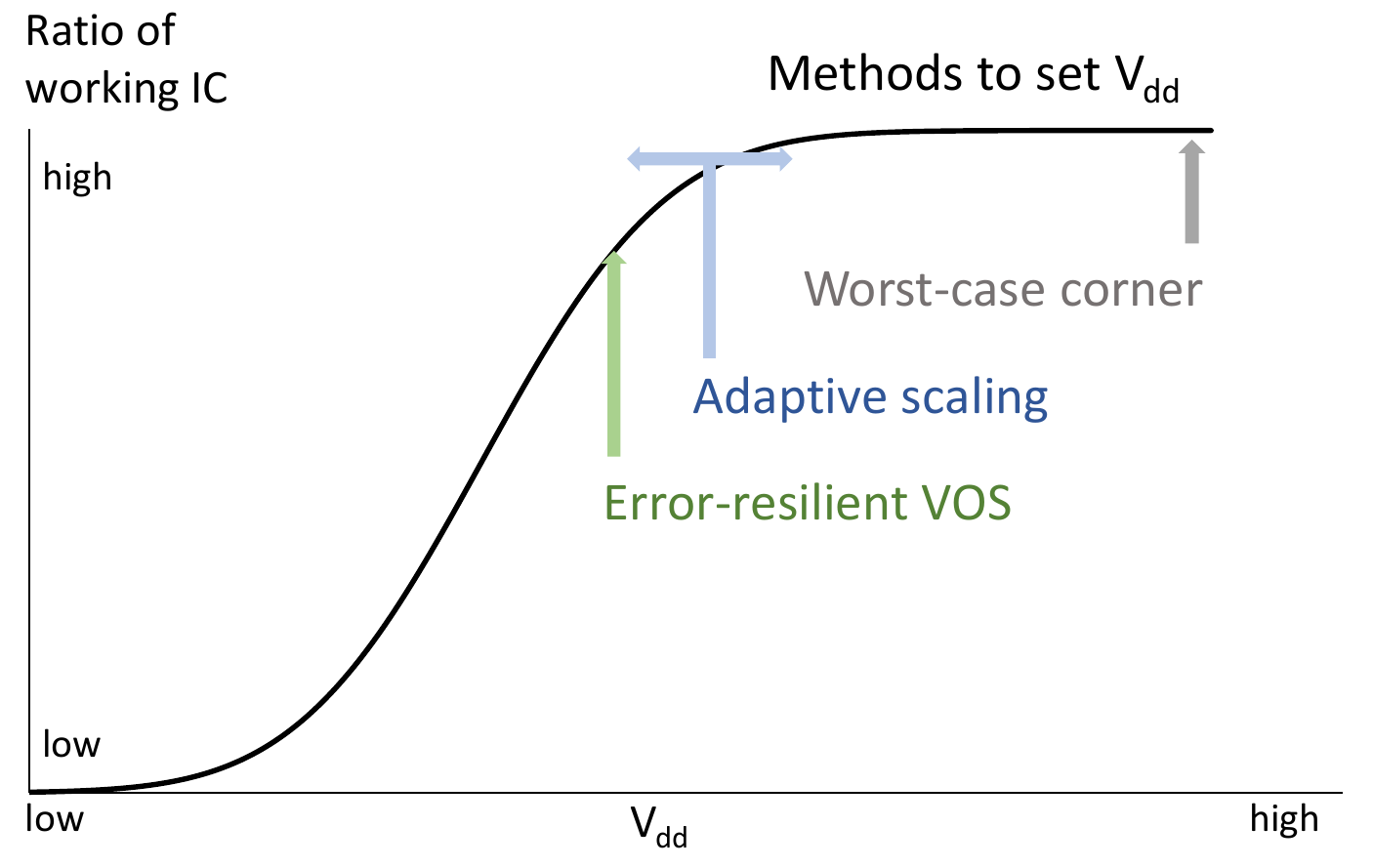}
\caption{Different approaches to scaling of the supply voltage $V_\text{dd}$ to cope
with speed variability. Operation at the worst-case corner misses out
on potential energy savings. Adaptive Voltage Scaling (AVS) provides
the just-needed $V_\text{dd}$ for the circuit to function error-free. A
further reduction of $V_\text{dd}$ by Voltage Over-Scaling (VOS) would save
more power, yet would introduce processing errors.}\label{fig:VOS}
\end{figure}

However with scaling towards deep sub-micron CMOS technologies (65~nm
and smaller), designers are facing ever-increasing variability
challenges. The process, voltage and temperature (PVT) variabilities
are considered to be the three main contributors to circuit
variability. Conventionally, to cope with this challenge, ICs are
designed at the worst PVT corners, to ensure that they always operate
correctly. Figure~\ref{fig:VOS} illustrates the different operating
regions for ICs suffering from manufacturing variability.

The conventional design approach for worst-case conditions introduces
considerable margins, leading to reduced peak performance and wasted
power consumption. The worst-case synthesis assumes that all devices
in the circuit operate in the slow-process corner and experience the
least favorable voltage and temperature conditions. Temperature
variations can yield up to 20\% speed differences for a single D
flip-flop. For instance, \cite{Huang2016} shows that for 28-nm
technology, the performance (speed) difference for a representative circuit is as large as $2.2$
times between the typical case and the worst case.  Adaptive scaling
techniques manage power dissipation and temperature by using a
variable supply voltage $V_\text{dd}$.

Scaling down the supply voltage is regarded as an error-free power
saving method as long as the signal timing constraints are
met. 
However, the critical (minimum) $V_{dd}$ that
guarantees timing closure cannot be determined at design time due to
PVT variabilities and aging effects.

A third design approach has recently gained interest, namely to scale
the $V_\text{dd}$ below the critical supply voltage, which is called
Voltage Over-Scaling (VOS).  In the VOS approach, the designer accepts
that sporadic errors might occur: for logic components, the signal
from the longest propagation paths can be mis-captured; for memory
components, it may lead to incorrect write/read data/address or data
loss.  This methodology of approximate computing enables very
energy-efficient processing \cite{Han2013}.
Wireless communication systems are designed to cope with
distortions and errors occurring on the channel. They are hence
inherently good candidates for error-resilient processing
solutions. 
In Massive MIMO, the large
number of antennas implies redundancy in the system. It is promising
to apply VOS specifically in the per-antenna processing,
reaching beyond the reliability margins of the circuits, but still
operating at a point where the computations are more often correct
than wrong.

\subsection{Massive MIMO Resilience to Circuit Errors}

Massive MIMO inherently is resilient to some circuits errors in the
per-antenna processing. Hardware errors in a number of antenna paths
can be absorbed by the system thanks to the averaging induced by the
large number of antennas -- reminiscent of how the effects of small-scale fading
 average out in the coherent multi-user MIMO
processing \cite{Marzetta16book}. Semiconductor process variability was at first experienced globally, between wafers or circuits separated in space on a silicon wafer, hence die-to-die. Designers have thus realistically assumed transistor parameters to be correlated for nearby circuits on a specific die and chip.  However, in deeply scaled technologies, device variability is mostly caused by the inaccuracy of lithography and etch technology. Intra-die (local) variations have consequently become significant, and are even reported dominant over global variations \cite{Saha2010}. This apparent design challenge comes with a new opportunity to shave margins in the implementation of Massive MIMO. Indeed, different
from the distortion resulting from non-linearities, the digital
distortion is independent of the signal and hence uncorrelated over
the antennas. The massive MIMO system will continue functioning even
when, sporadically, one or a few individual antenna signals is subject
to full failure. As mentioned in Section~\ref{sec:edge}, this opens
the door to operation of circuits with much lower design margins
compared to traditional specifications, and most interestingly at
lower supply voltages and hence power consumption.

\begin{figure}[t!]\centering
	\includegraphics[width=1\linewidth]{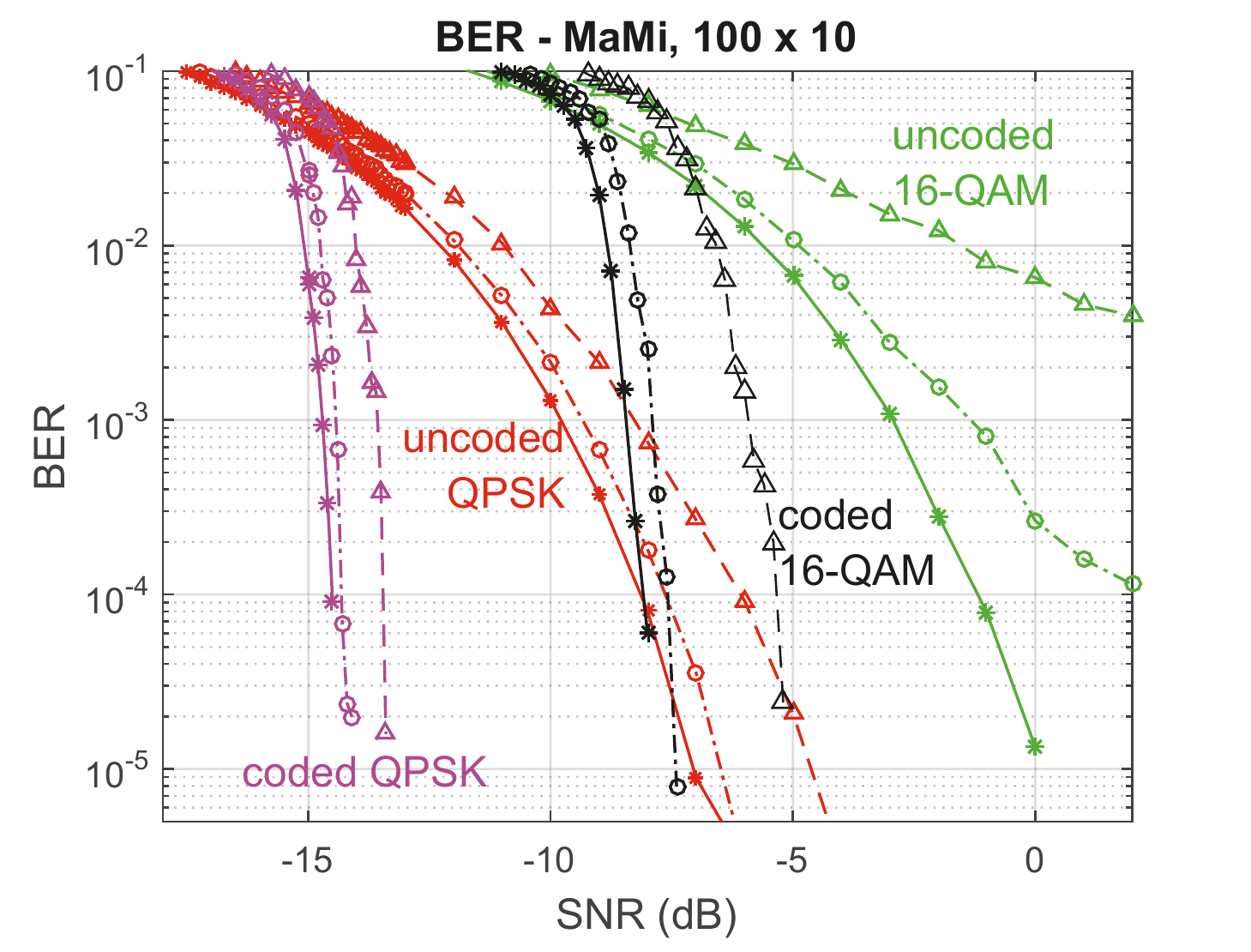} \caption{BER
	performance versus channel SNR. Randomly affected ``victim
	antennas"  from significant digital hardware errors for uncoded and coded (3/4 soft LDPC) QPSK, and uncoded and coded (3/4 soft LDPC) 16-QAM. From \cite{Huang2017}. The legend denotes: i) error-free (star markers), ii) 3\% victim antennas (circle markers), and iii) 10\% victim antennas (triangle markers).}\label{fig:SDDR}
\end{figure}

The digital hardware errors in (I)FFT and filters introduced by silicon
unreliability and by ambitious design methodologies result in
incorrect bits during signal processing. This can be regarded as
digital circuit distortion. We characterize the impact on the purity
of the signal in terms of  the signal-to-digital distortion ratio (SDDR):
\begin{equation}
\text{SDDR} = 10 \cdot \log \frac{{\sigma_s}^2}{ {\sigma_d}^2 }
\end{equation} 
where ${\sigma_s}^2$ and ${\sigma_d}^2$ are the powers of the
error-free digital antenna signal output, and the noise power of the
digital distortion due to circuit unreliability, respectively. First,
we consider VOS errors which are temporary and local in nature. The
BER-performance is shown in Figure~\ref{fig:SDDR} for a severe SDDR distortion, where signals get stuck at a fixed value. Results for different modulation orders and both uncoded and coded performance (rate 3/4 soft decoded LPDC) are shown. The resulting SNR degradation remains limited to $<1$~dB for $3\%$ of the antennas being a ``victim'' of circuit errors in the coded 16-QAM, and even up to $10\%$ of the antennas in QPSK case. 

\begin{figure}[t!]\centering
	\includegraphics[width=1\linewidth]{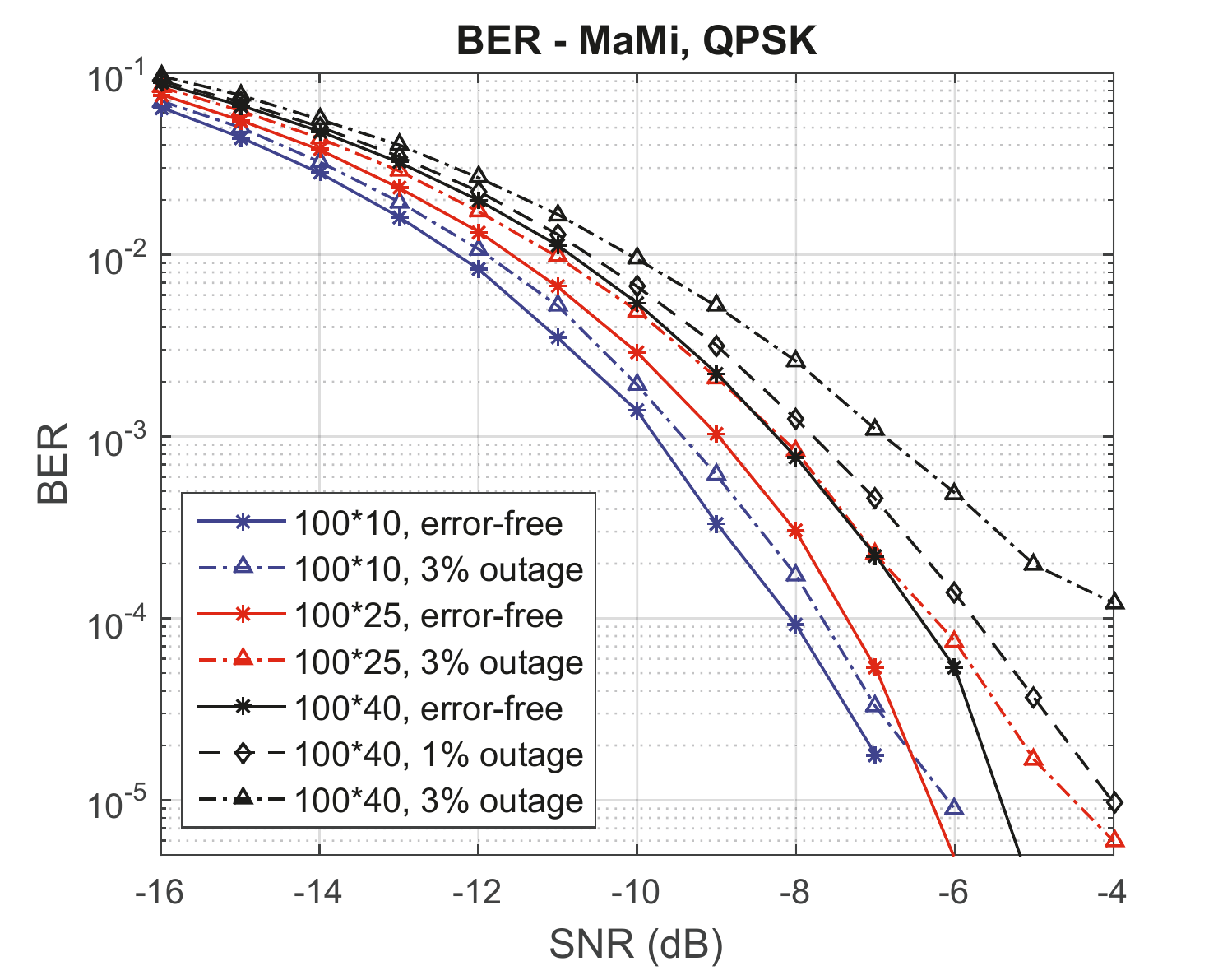} \caption{Impact
	of antenna outage on Massive MIMO system performance depends
	on the system load, for the pessimistic case where the errors
	are not detected. Disabling antennas will limit the impact of
	antenna outage on the Massive MIMO system
	performance. From \cite{Huang2017}.}\label{fig:fail}
\end{figure}

When operating deeply scaled circuits without margins, occasionally a
full circuit failure may occur. The impact of this effect on the
Massive MIMO system performance is called ``antenna outage''. The
digital output are then permanently stuck at a fixed value, which
is assumed to be its maximum possible value. The SDDR of the outage
antenna is $-\infty$, as the signals from the victim antennas are
completely lost. This model is regarded a worst-case hardware
failure. Note that the $-\infty$ SDDR does not imply infinite noise to
the whole system, as only the victim antennas are affected and their
PA power is normalized among all antennas. Therefore, a single antenna
outage will not cause the system to fail entirely. The impact on the
system performance is shown in Figure~\ref{fig:fail} for different
antenna outage and system loads, for the pessimistic case where the
errors are not detected.

As demonstrated, Massive MIMO can operate well with rather severe
circuits errors, and thus allows significant VOS. The
impact increases with higher system load and modulation
constellations.
 The $V_\text{DD}$ may be adapted according to the system
parameters to always offer just sufficient performance. In-situ
monitoring based on test signals can be applied to perform adequate
$V_\text{DD}$ scaling \cite{Huang2017}.

\begin{figure}[t!]\centering
	\includegraphics[width=0.9\linewidth]{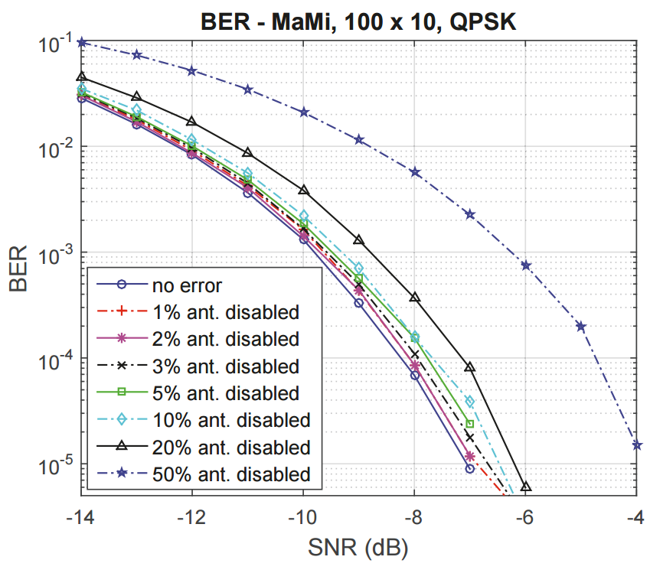} \caption{BER
		performance is only slightly degraded for up to $\sim 10\%$ of antennas failing. Systematic failure of circuits is detected and corresponding antenna signals are discarded. From \cite{Huang2017}.}\label{fig:discarded}
\end{figure}

In order to further improve the system robustness towards hardware
errors, techniques to first detect hardware errors, and next either
neglect, or if needed disable, defective hardware can be
applied. Importantly, the distortion originating from digital circuit
errors fundamentally differs from pure random noise. While process
variations may feature continuous random distributions, their effect
typically results in discrete error events. Dedicated monitoring
circuitry can be established \cite{fojtik2012}
for the functional components such as (I)FFTs and filters, that will detect these errors. If the Massive
MIMO system is operated whereby it receives information from the hardware
level on failing circuits, it can adapt its signal processing accordingly. One option is
to disable systematically failing antenna paths and no longer consider
them in the central processing. The BER results are given in Figure~\ref{fig:discarded} for a case with moderate system load (10$\times$100 in this simulation). It shows that   excluding defective circuits limits the degradation level to $<0.5$~dB on uncoded QPSK for up to $\sim 10\%$ of the antenna paths failing. This approach is equivalent to
operating the Massive MIMO system with a reduced number of BS antennas
$M$. For a representative case of QPSK transmission in a 100-antenna, 10-user scenario and with
28~nm standard CMOS technology, up to 40\% power savings can be
achieved with negligible performance degradation \cite{liu2010}. 

In conclusion, lean   per-antenna processing can
be performed in Massive MIMO systems. The very large number of
operations, due to the large number of antenna paths, can be
performed with low precision and with a profoundly scaled supply
voltage. Combined, these techniques can reduce the power consumption due
to the digital processing on each antenna path by an order of
magnitude. For an exemplary system with 100 antennas at the base
station, the total is comparable to a conventional MIMO system with an
order of magnitude less antennas.

%% file: terminal.tex
\subsection{Increased Service Levels on Low Complexity Terminals}

It has been shown that the Massive MIMO system concept does
not require any additional specific functionality at the UE side. 
Massive MIMO terminals that have a single antenna, or apply simple diversity reception, will only be able
to receive a single spatial stream. However, large numbers of terminals can be multiplexed in the same
time-frequency slot, and every terminal can be allocated the full bandwidth of the system.
This results in a  throughput per terminal  comparable with that  of conventional UEs
 that receive multiple spatial streams in  parallel.

\begin{figure}[t!]
	\centering
	\includegraphics[width=0.5\textwidth]{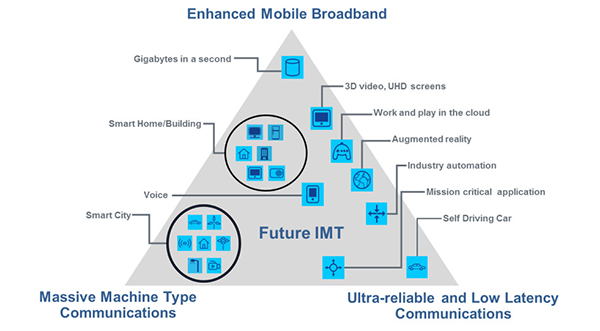}
	\caption{Envisioned use cases for future international mobile
          telecommunication. (source: Recommendation ITU-R M.2083-0 ``Framework and overall objectives of the future development of IMT for 2020 and beyond'' \cite{ITU_5G})} \label{fig:5Gtriangle}
\end{figure}

5G terminals are expected to come in large numbers and support a
diverse set of service requirements. Next to the continued traffic
growth towards terminals allocated to human users, a variety of
devices will require Machine Type Communication
(MTC). Figure~\ref{fig:5Gtriangle} illustrates three main use cases
 envisioned by
industry alliances and the International Telecommunication Union (ITU) \cite{ITU_5G}.

Figure~\ref{fig:5Gtriangle} demonstrates that 5G technologies not
only need to enhance mobile broadband links. New solutions are
needed to connect a very large number of (ultra-) low-power devices and
machines requiring very reliable and low-latency services. Massive
MIMO can simultaneously support many broadband terminals in
sub-6~GHz bands in indoor, outdoor, and mobile environments. The technology
can also be tailored to optimally serve new MTC-based applications.
Especially for narrowband MTC, the  high array gain and the high degree of spatial diversity offered by Massive MIMO 
will help. The spatial diversity specifically gives rise to channel hardening.  

\begin{figure}[t!]
\centering
\includegraphics[width=0.4\textwidth]{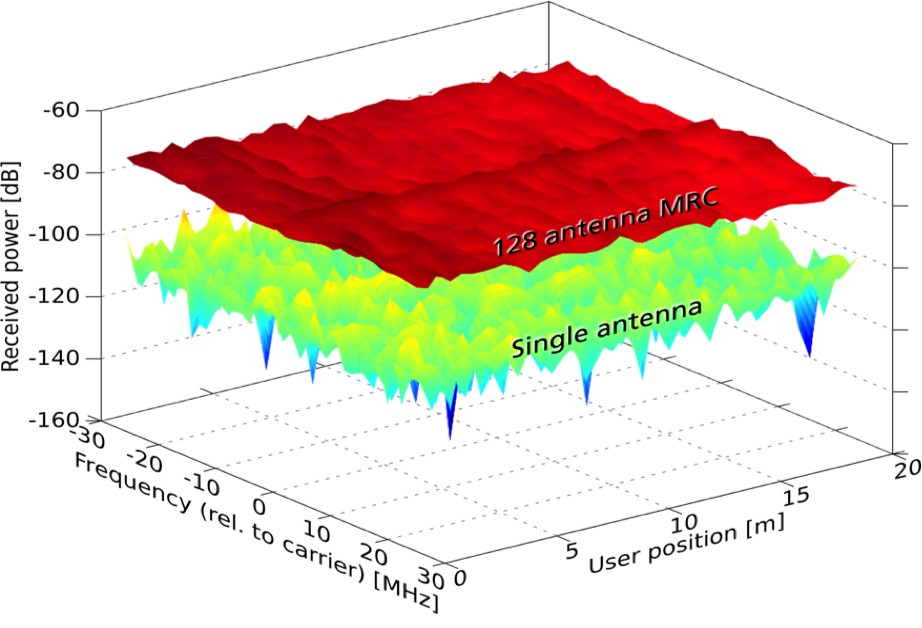}
  \caption{The array gain and channel hardening effect demonstrated
    experimentally, for a $M=128$, $K=8$ setup. With permission and
    \copyright Ove Edfors, Lund University. \label{fig:channel_hard}}
\end{figure}

The effects of   array gain and channel hardening are   
illustrated for a 128-antenna setup  in Figure~\ref{fig:channel_hard}.
Consistently boosted signal levels  over all terminal positions, thanks to the array gain, are
observed. Terminals can potentially transmit data at  several
tens of dB
lower output powers. The latter however requires high-quality CSI to be available, and the
power allocated to  pilots will limit the savings in practice. The channel hardening effect enhances the
reliability of the links and improves the quality of service; most
notably:
\begin{enumerate}
\item Increased performance at the cell edges, where terminals may experience limited or worst case no connectivity in current networks. Massive MIMO addresses
  this challenge, provided good uplink pilot-based CSI acquisition is
  ensured.

\item Power savings and hence longer autonomy for battery-powered
  devices.

\item Improved reliability. Fewer packet retransmissions can also
  reduce the end-to-end latency. The specifications put forward for Ultra
  Reliable Low Latency Communication (URLLC) in 5G is to support a
  $99.9999\%$ reliability, and an end-to-end latency better than 1ms.

\item Sustained good service levels in conditions with many
  simultaneously active users.
\end{enumerate}

In the next paragraphs, first a typical broadband user equipment is
zoomed in on. It is indicated how low-power operation can be
achieved while keeping backward compatibility with 4G air
interfaces. Next, we discuss how tailored Massive-MIMO systems have great
potential to address the challenging requirements of MTC terminals.

\subsection{Energy Efficient Broadband Terminals}

\begin{figure}[t!]
\centering
\includegraphics[width=0.5\textwidth]{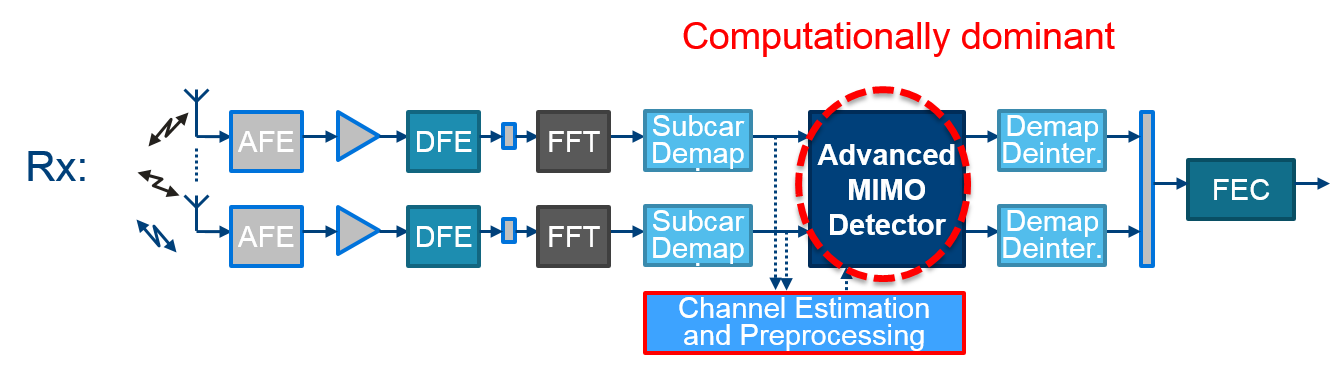}
  \caption{A conventional wideband receiver for multiple spatial
    layers requires complex MIMO detection. \label{fig:ConvRx}}
\end{figure}

No advanced processing is required at the UE in Massive MIMO systems. 
In contrast, 4G systems deliver broadband services to UEs
through multiplexing of several spatial layers. We compare a typical Massive MIMO terminal with the reference case of a $4\times4$ MIMO link. The latter requires MIMO
detection at the terminal side in the downlink.  Figure~\ref{fig:ConvRx}
shows a functional block scheme of a conventional broadband, multiple-antenna terminal receiver.

The complexity breakdown of a typical MIMO-OFDM baseband chain
identifies   channel estimation  and MIMO
detection as the main bottlenecks. We take as a reference $4\times
4$ MIMO-OFDM case where the multiple-antenna processing can be
conveniently performed per subcarrier resulting in relatively low-complexity
 implementations \cite{Vandenameele2000}. We consider a
basic linear MIMO detector, and non-linear detectors implementing (ordered) Successive
 Interference Cancellation (SIC). The latter are
required to achieve acceptable system performance especially in the
low-SNR regime and in high-mobility scenarios. The power
consumption of the inner modem receiver of the terminal in a Massive
MIMO system is estimated relatively to published VLSI implementations for
conventional MIMO receivers \cite{Yoshizawa2009, Ketonen2010}. A range
of algorithms and implementations for MIMO detectors have been
reported on, differing substantially in complexity.  Our analysis is
based on typical data for the specific components, and our own design
know-how. Table~\ref{tb:UEcomplexity} summarizes the assessment for
both single-antenna and dual-antenna diversity-reception terminals,
demonstrating an expected reduction in power consumption of a factor 5 to
50.\footnote{A similar reduction in hardware complexity could be
  achieved for UE radios custom-designed to operate in Massive MIMO
  networks specifically. Backward compatibility with previous
  broadband systems may require the presence of MIMO detection
  hardware in broadband UEs in practice.}
The instantaneous throughput will be higher for conventional MIMO terminals receiving several spatial layers. To compare the energy efficiency (in Joule/bit), the same average throughput needs to be considered.    

\begin{table}[t!]
	\caption{Relative power consumption estimates for UE inner modem receivers.}
	\label{tb:UEcomplexity} \centering
	\begin{tabular}{ | c | c | c | c | }
		\hline
		\textbf{\textit{4 x 4}}  & \textbf{\textit{4 x 4}}
		& \textbf{\textit{Massive MIMO}}  & \textbf{\textit{Massive MIMO}} \\
		
		\textbf{\textit{linear}}  & \textbf{\textit{non-linear}}
                        & \textbf{\textit{}}  & \textbf{\textit{2-antenna}}  \\
        	\textbf{\textit{detector}}  & \textbf{\textit{detector}}
        & \textbf{\textit{single antenna}}  & \textbf{\textit{diversity}}
		\\\hline
		$P_{ref}$ &	$1.5 - 5  P_{ref}$	& $\sim 10\% P_{ref}$ &  $\sim 20\% P_{ref}$	\\\hline
	\end{tabular}
\end{table}

\subsection{Tailored Solutions Fit for Low-Power Connected Devices}

\begin{figure}[t!]
	\centering
	\includegraphics[width=1\linewidth]{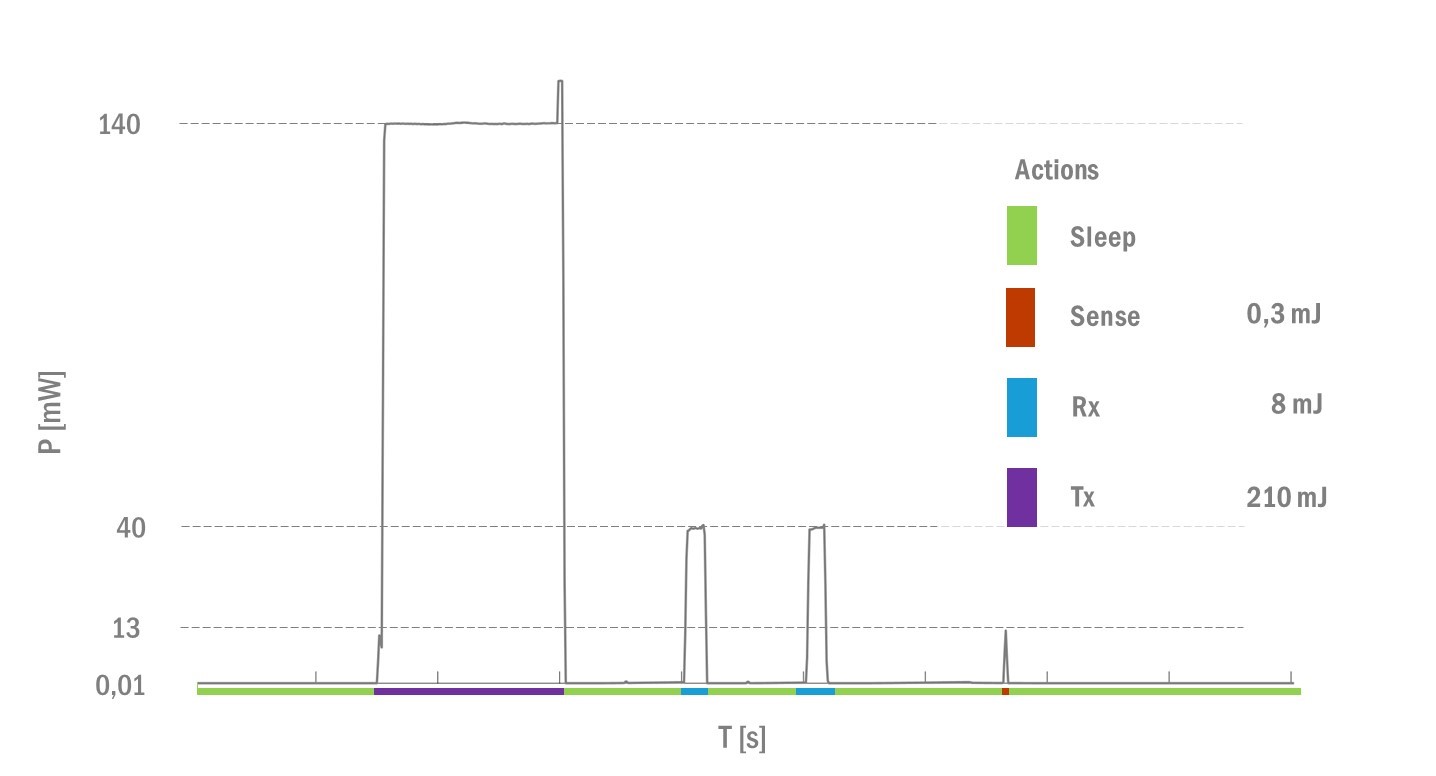}
	\caption{The transmit energy will dominate the battery time on a LPWAN IoT node.}\label{fig:LoraEnergy}
	\label{fig:loraEnergy}
\end{figure}

MTC for sensors and actuators opens the door for a variety of new IoT
applications.  Low energy consumption is essential to enable long
autonomy of devices powered by batteries or even relying on harvested
energy. The physics of radio propagation dictates a strong attenuation
on the link with distance, $d$:
\begin{equation}
P_\text{Rx} \propto G_\text{Tx}G_\text{Rx}d^{-n}P_\text{Tx},\ n=2\ \mbox{in\ free\ space},  n > 2\ \mbox{typically},
\label{eq:PathLoss}
\end{equation}
where $P_\text{Rx}$ and $P_\text{Tx}$ are the received and transmitted powers,
respectively, and $G_\text{Rx}$ and $G_\text{Tx}$ are directivity gains at
the receiving and transmitting end of the link.  The above is
especially unfortunate for mostly uplink-dominated MTC.  Low Power
Wide Area Network (LPWAN) technologies are dedicated to connect IoT
nodes at long ranges. We performed   measurements with an IoT node communicating via a LORA gateway \cite{DRAMCO_tutorial}.  Inspection of the power consumption of this
illustrative node in Table~\ref{tb:LoraPower} provides valuable
insights. The transmit power is relatively
high since the power amplifier needs to provide sufficient power to
cope with large-scale fading losses. The energy consumption, which
will ultimately determine the autonomy of the node, is shown in
Figure~\ref{fig:LoraEnergy}.

\begin{table}[t!]
\caption{Power consumption in different modes measured on a LPWAN IoT node.}
\label{tb:LoraPower} \centering
\begin{tabular}{ | c | c | }
\hline
\textbf{\textit{Operation mode}}  & \textbf{\textit{Power consumption (mW)}} \\ 
\hline
Transmit  &	$\ge 140$\footnote{The module used to perform the measurements has a current limited to 45~mA.}	 \\\hline
Receive &	$40$		\\\hline
Sense & $13$	\\\hline
Sleep &  $0.1$	\\\hline
\end{tabular}
\end{table}

This pinpoints the fierce challenge of connecting sensor nodes and
other autonomous devices at a longer range. Their traffic is mostly
dominated by uplink, hence putting the node in the most energy-consuming transmitting mode. Equation~(\ref{eq:PathLoss}) reveals that
fundamentally only few parameters can be influenced to improve the
link budget. Antennas at IoT nodes, due to size and cost constraints,
 can hardly offer any gain and on the contrary not seldom introduce losses.  Massive MIMO systems exhibit a
large-antenna array gain and apply an adaptive channel-matched
beamforming approach. They offer the opportunity to reduce the
transmit power in constrained MTC nodes proportionally to the square
root of number of BS antennas $M$ or even proportionally to $M$ if
accurate CSI is acquired. This enables the simultaneous service of a large number of
 devices. This asset
is important to keep up with the predicted evolution towards Massive
MTC.  A Massive MIMO-based LPWAN could also offer extended
coverage and increased reliability, provided that a power-efficient solution for the pilot-based
CSI-acquisition is implemented. This challenge, to develop Massive
MIMO technology for MTC services is further discussed in
Section~\ref{sec:conclusions}.

%% file: conclusions.tex

\subsection{Signal Processing at Work in Massive MIMO Demonstrations}

Demonstrations that have proven the superior
spectral efficiency of Massive MIMO and the adequacy of DSP solutions
in real-life testbeds are illustrated here below. Furthermore we summarize the conclusions of this
paper and outline future research directions.

\begin{figure}[t!]
    \centering
    \begin{subfigure}[]
        \centering
        \includegraphics[width=0.20\textwidth]{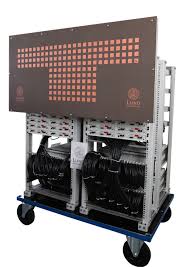}
    \end{subfigure}
    \begin{subfigure}[]
        \centering
        \includegraphics[width=0.25\textwidth]{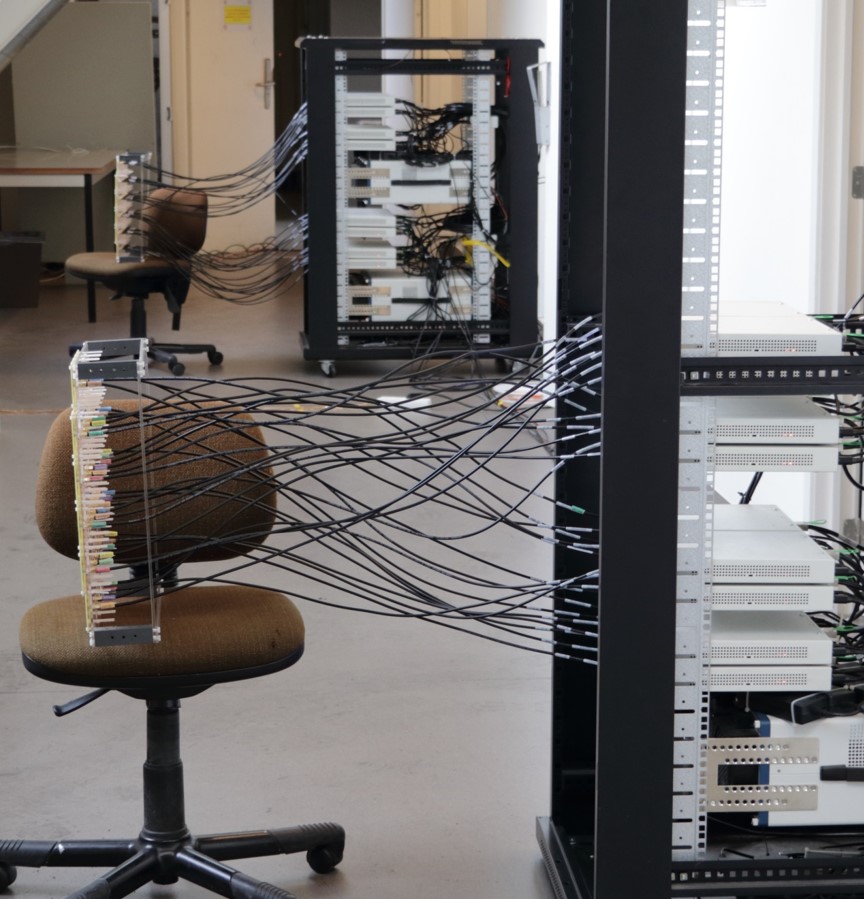}
    \end{subfigure}
    \caption{Two different Massive MIMO testbeds: (a) the LuMaMi testbed at
    Lund University a with collocated antenna array (from \cite{Malkowsky2017IEEEACESS}) and (b) the KU Leuven testbed with separated antenna arrays.}
\label{fig:lumami}
\end{figure}

To prove a new wireless technology, it is very important
to build up testbeds to conduct verification and evaluate performance
in real-life environments with over-the-air transmission. For Massive
MIMO it is especially crucial, since  performance is
dependent on propagation characteristics, and measurement-based
channel models  themselves are still under development. Thanks to 
recent advances in Software-Defined Radio (SDR) technology, several
Massive MIMO prototype systems have been built by both industry and
academia, including the Argos testbed with 96 antennas \cite{SYA2012},
Eurecom's 64-antenna testbed \cite{TestbedEurecom}, Facebook's ARIES
project \cite{TestbedFacebook}, the 100-antenna LuMaMi testbed from
Lund University
(Figure~\ref{fig:lumami}a) \cite{Malkowsky2017IEEEACESS}, SEU's
128-antenna testbed \cite{SEU_NUPT}, and testbeds exploring
distributed arrays from the KU Leuven
(Figure~\ref{fig:lumami}b) \cite{Chen2016} and University of
Bristol \cite{Harris2016SIPS}.

 \subsubsection{World-Record in Spectral Efficiency and Massive MIMO in Mobility}

The signal processing techniques discussed in this paper, especially
the cross-level optimization methodology, have been exploited in the
development of Massive MIMO testbeds to enable real-time processing of
wide-band signals for large numbers of antennas. For instance, the
LuMaMi tested adopts the processing distribution scheme   in
Figure~\ref{fig:proc_dist}, where 50 SDRs with Field-Programmable
Gate-Arrays (FPGAs) are used to perform per-antenna processing in a
parallel fashion. Four centralized FPGAs are responsible for
per-subcarrier processing, and the Peripheral Component Interconnect
Express (PCIe) with direct memory access (DMA) channels handles the
 data shuffling. QR-decomposition based ZF processing has
been implemented to fully leverage the available parallel processing
resources in the FPGAs.

Diverse field trials, both indoors and outdoors with static and mobile
users, have been conducted using the Massive MIMO testbeds. In a 2016
experiment, a 128-antenna Massive MIMO base
station   served 22 users, each transmitting with 256-QAM
modulation, on the same time-frequency
resource \cite{Harris2016SIPS}. The spectral efficiency benefits from
the spatial multiplexing as well as from the high constellation order,
enabled by the array gain.  In practice, protocol overhead and FEC
redundancy will determine the actual net spectral efficiency.  In the
actual demonstration a spectral efficiency of 145.6 bits/s/Hz was
achieved on a 20~MHz radio channel, representing a $\sim 20$~times
increase with respect to the current 4G air interface. The performance
was achieved in an environment without mobility and multi-cell
interference, which would constitute the limiting factors  
performance in a practical deployment.

\begin{figure}[t!]
\centering
\includegraphics[width=0.45\textwidth]{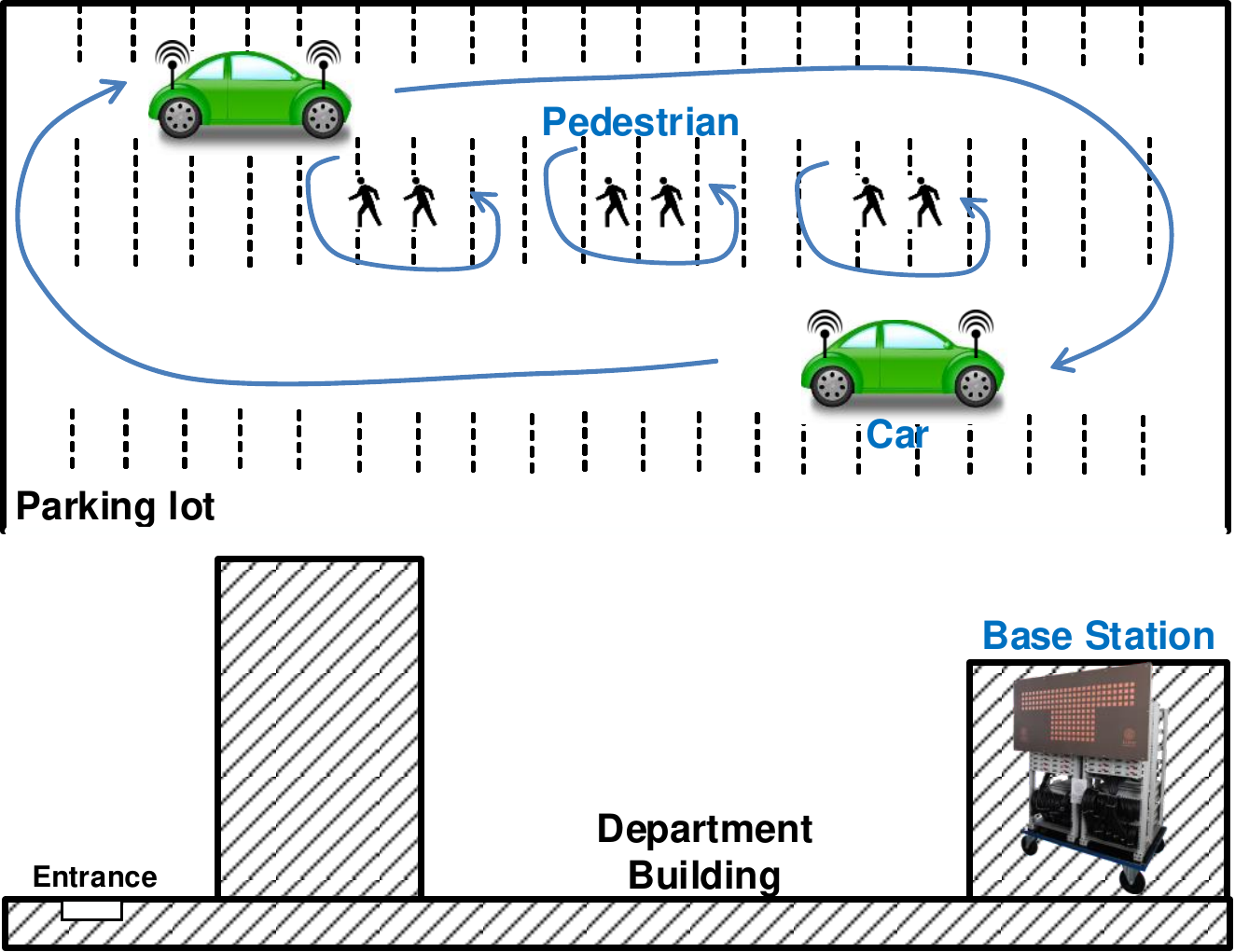}
\caption{Overview of the testbed demonstration of Massive MIMO in a mobility scenario, at the campus of Lund University, Sweden.}
\label{fig:moblity_test}
\end{figure}

The same research group also demonstrated Massive MIMO operation in an
outdoor scenario with moderate
mobility \cite{Harris2017JSAC}. Figure~\ref{fig:moblity_test} shows
the measurement scenario where the 100-antenna LuMaMi testbed is
placed on the rooftop of a building facing a parking lot $\sim 75$~m
away. Ten single-antenna users are served in real time at 3.7~GHz,
including six users moving at pedestrian speed and four terminals on
vehicles moving at a speed up to around $50$ km/h. The spatial
multiplexing was fully achieved and the communication quality was on
average well maintained for all terminals \cite{MAMMOETD4_2}. Sporadic
interruptions could be traced back to temporary loss of
synchronization. It should be noted that both the speed of the cars
and the number of terminals could be larger in a real deployment. In
the proof of concept they were limited by the available test space and
equipment. In fact, at 3.7 GHz carrier frequency and with a slot
length of 0.5~ms, the maximum permitted mobility (assuming a two-ray
model with Nyquist sampling, and a factor-of-two design margin, as
in \cite{Marzetta16book}) is over 140 km/h \cite{MaMiBlog2016}.

\subsubsection{Further Investigation Needed for Synchronization}

A critical challenge requiring further investigation is the initial
synchronization between the base station and the user
terminals. This initial synchronization has to start without
any knowledge of the channels, and therefore cannot  
benefit from an array gain. How to efficiently perform initial
time and frequency synchronization acquisition without the massive
array gain and how to explore the (partial) array gain to provide
faster and more robust synchronization are still open 
questions. Two methods were studied during the LuMaMi
testbed experiments. One method is to reserve a dedicated RF chain for
the synchronization signal, which is transmitted using an
omni-directional antenna. In this case,  a higher-power
 PA (which is not available in LuMaMi) is needed to
provide  coverage. Another method is to use  
beam-sweeping for the synchronization signal \cite{Barati2015TWC}, but
this is inefficient, as it is essentially equivalent to  repetition
coding, and also there is risk of synchronization loss when the users
are not hit by a beam.  Improved techniques, based on space-time block
codes, have been investigated  
\cite{Karlsson2017arXiv,Xia16,Meng16}. Iterative search and
tracking methods \cite{Marco2016CM} may have potential,
especially for mobile users.

\subsection{Concluding on the Signal Processing}

Appropriate co-design of algorithms, hardware architectures, and
circuits in Massive MIMO implementations brings significant benefits:
\begin{itemize}
\item Energy efficient implementations of ``theoretically optimal''
  Massive MIMO DSP architectures are nontrivial but possible. We have
  detailed some of the most important innovations required, and
  explained their analysis. The power consumption of conventional
  macro base stations is dominated by the PA stage. They benefit in
  Massive MIMO from the ability to operate on an order of magnitude
  less transmit power.

\item The sufficiency of low-precision quantization and processing,
  predicted by information-theoretic studies, has now also been
  validated through real signal processing experiments. A reduction in
  word-length up to 6 times compared to conventional systems
  translates into corresponding savings in complexity, power
  consumption and memory.

\item Dedicated and scalable hardware architectures implementing tailored
  algorithms for large matrix processing facilitate zero-forcing
  precoding at the base station in real time, at 30 mW power
  consumption in relevant scenarios for a $128\times 8$ system.

\item Voltage over-scaling, a speculative concept just 5 years ago, has
  found   appropriate application in the Massive MIMO per-antenna
  processing.

\item Smart control of algorithmic modes and scalable devices,
  including body bias adaptation, can guarantee suitable
  performance-power trade-offs over a wide range of communication
  scenarios and channel propagation conditions.

\item Lean terminals could operate in typical broadband cellular Massive MIMO
 networks at about $10\%-20\%$ of the power consumption 
of equivalent conventional terminals, both in data
  transmission and reception.
  
 \item The efficiency of Massive MIMO base stations can be further
 improved by relaxing the requirements of the RF and analogue
 hardware. However, caution is needed as (non-linear) distortion may
 under specific conditions combine coherently.
 
\end{itemize}

\subsection{Future Directions}

\subsubsection{Progress Massive MIMO Deployment in Actual Networks}

Integration of  all   components into    deployment in actual networks
 represents a vast design and
development effort, that will include:
\begin{itemize}
\item Overcoming challenges related to connection of the many antenna paths
to the central processing units. This involves implementing high-speed
interconnects and coping with potential coupling effects in the
front-end modules.

\item Devising efficient schedulers for large numbers of users. 
Achieving the high spatial multiplexing 
gains offered by Massive MIMO fundamentally requires that many terminals are
scheduled for service simultaneously.  Tuning or re-design of higher-layer protocols
  could be beneficial to shape  the traffic patterns, such that aggressive spatial
  multiplexing can be performed.

\item Designing   antenna arrays.  
Massive MIMO arrays do not have to be linear, rectangular or cylindrical.
Small antenna elements could be naturally integrated into the environment,
 onto the surface of existing structures, or faces of buildings, for example,
  in an aesthetically pleasing manner.

Insights from electromagnetics may guide the design of new types of arrays.
Specifically,  for a given volume $V$, consider the corresponding
  smallest possible sphere that contains $V$. If one covers the
  surface of this sphere with antennas at a density of $\sim
  1/\lambda^2$ elements per square meter, then there is no point in
  installing any additional elements inside of the interior of $V$
  \cite{FranceschettiM2015}. Sampling the surface on a
  $\lambda\times \lambda$-grid captures all information in the
    radiated field. In conclusion, what goes into the interior of $V$ is unimportant,
    only the surface matters.  
\end{itemize}

Industrial recognition of the value of Massive MIMO technology is
evidenced by the large number of contributions on the topic in the
3GPP-LTE standardization of New Radio (NR) for 5G systems. Leading
operators have already started to perform commercial field trials of the technology
\cite{MaMiBlog2016}.  

\subsubsection{Enhanced Functionalities} 

Large antenna arrays can also be used to perform accurate positioning
and localization. This feature can offer improved context-awareness to
services. Also the Massive MIMO communication system itself could
exploit this information to perform smart pilot allocation, for example.

\subsubsection{Scale Up Capacity and Efficiency}

The call for more and higher-quality wireless services is expected to
increase for many years, and the quest for wireless systems offering
higher spectral and energy efficiency will continue. Higher peak-rates
can be offered in Massive MIMO by performing spatial multiplexing of
several streams to one terminal. Actual gains may be limited due to
insufficient rank of the channel, yet for two streams this will mostly
be achievable with co-located antennas exploiting cross-polarization.

Wider bandwidth channels can be allocated especially in mmWave bands. Radio propagation and in particular absorption is considerably different at these frequencies. Arrays with a large number of antennas can be small in size, yet their effective gain may suffer from high losses on the interconnect. Consequently, Massive MIMO systems in these bands call for other architectures and their deployment will best suit particular use cases, for example hotspots.

With larger antenna arrays, both  better spatial multiplexing and
array gains  can be achieved. The new concepts of cell-free Massive MIMO \cite{ngo2017cell} 
and intelligent
surfaces \cite{LIS2017} accelerate this trend to a next level. 
With cell-free Massive MIMO, coherently cooperating antennas are spread out over a larger geographical area, providing improved macro-diversity and improved channel rank for
multiple-antenna terminals.
The intelligent surface concept  envisages
distributed nodes that form electromagnetically active walls, floors,
and planar objects.    
New research is urgently needed to bring these new concepts to their full potential.

%% file: SP_overview_final.bbl

%% file: SP_overview_arxiv.bbl
\begin{thebibliography}{10}
\providecommand{\url}[1]{#1}
\csname url@samestyle\endcsname
\providecommand{\newblock}{\relax}
\providecommand{\bibinfo}[2]{#2}
\providecommand{\BIBentrySTDinterwordspacing}{\spaceskip=0pt\relax}
\providecommand{\BIBentryALTinterwordstretchfactor}{4}
\providecommand{\BIBentryALTinterwordspacing}{\spaceskip=\fontdimen2\font plus
\BIBentryALTinterwordstretchfactor\fontdimen3\font minus
  \fontdimen4\font\relax}
\providecommand{\BIBforeignlanguage}[2]{{%
\expandafter\ifx\csname l@#1\endcsname\relax
\typeout{** WARNING: IEEEtran.bst: No hyphenation pattern has been}%
\typeout{** loaded for the language `#1'. Using the pattern for}%
\typeout{** the default language instead.}%
\else
\language=\csname l@#1\endcsname
\fi
#2}}
\providecommand{\BIBdecl}{\relax}
\BIBdecl

\bibitem{nr}
``First {5G NR} specs approved, {Dec. 2017},''
  \url{http://www.3gpp.org/news-events/3gpp-news/1929-nsa\_nr\_5g}.

\bibitem{Larsson2014}
E.~G. Larsson, O.~Edfors, F.~Tufvesson, and T.~L. Marzetta, ``Massive {MIMO}
  for next generation wireless systems,'' \emph{IEEE Comm. Mag.},
  vol.~52, no.~2, pp. 186--195, Feb. 2014.

\bibitem{Marzetta16book}
T.~L. Marzetta, E.~G. Larsson, H.~Yang, and H.~Q. Ngo, \emph{Fundamentals of
  {M}assive {MIMO}}.\hskip 1em plus 0.5em minus 0.4em\relax Cambridge
  University Press, 2016.

\bibitem{Harris2017JSAC}
P.~Harris, S.~Malkowsky, J.~Vieira, E.~Bengtsson, F.~Tufvesson, W.~Hasan,
  L.~Liu, M.~Beach, S.~Armour, and O.~Edfors, ``Performance characterization of
  a real-time {m}assive {MIMO} system with {LOS} mobile channels,'' \emph{IEEE
  Journal on Sel. Areas in Comm.}, vol.~35, pp. 1244--1253, Jun.
  2017.

\bibitem{SEU_NUPT}
X.~Yang, W.~Lu, N.~Wang, K.~Nieman, C.~K. Wen, C.~Zhang, S.~Jin, X.~Mu,
  I.~Wong, Y.~Huang, and X.~You, ``Design and implementation of a {TDD}-based
  128-antenna {massive MIMO} prototype system,'' \emph{China Communications},
  vol.~14, no.~12, pp. 162--187, Dec. 2017.

\bibitem{Molish2017}
A.~F. Molisch, V.~V. Ratnam, S.~Han, Z.~Li, S.~L.~H. Nguyen, L.~Li, and
  K.~Haneda, ``Hybrid beamforming for {massive MIMO}: A survey,'' \emph{IEEE
  Comm. Mag.}, vol.~55, no.~9, pp. 134--141, 2017.

\bibitem{bjornson2017massive}
E.~Bj{\"o}rnson, J.~Hoydis, L.~Sanguinetti \emph{et~al.}, ``Massive {MIMO}
  networks: Spectral, energy, and hardware efficiency,'' \emph{Foundations and
  Trends{\textregistered} in Signal Processing}, vol.~11, pp. 154--655, 2017.

\bibitem{li2017massive}
X.~Li, E.~Bj{\"o}rnson, E.~G. Larsson, S.~Zhou, and J.~Wang, ``Massive {MIMO}
  with multi-cell {MMSE} processing: exploiting all pilots for interference
  suppression,'' \emph{EURASIP Journal on Wireless Comm. and
  Netw.}, vol. 2017, no.~1, p. 117, 2017.

\bibitem{Steffen2016SIPS}
S.~Malkowsky, J.~Vieira, K.~Nieman, N.~Kundargi, I.~Wong, V.~Owall, O.~Edfors,
  F.~Tufvesson, and L.~Liu, ``Implementation of low-latency signal processing
  and data shuffling for {TDD} {m}assive {MIMO} systems,'' in \emph{Proc. of IEEE
  International Workshop on Signal Processing Systems (SiPS)}, pp.
  260--265, Dec. 2016.

\bibitem{SYA2012}
C.~Shepard, H.~Yu, N.~Anand, L.~E. Li, T.~L. Marzetta, R.~Yang, and L.~Zhong,
  ``Argos: Practical many-antenna base stations,'' in \emph{Proc. of ACM Int.
  Conf. on Mobile Computing and Networking (MobiCom)}, Istanbul, Turkey, Aug.
  2012.

\bibitem{DBLP:journals/pieee/PuglielliTLMLTW16}
A.~Puglielli, A.~Townley, G.~LaCaille, V.~Milovanovic, P.~Lu, K.~Trotskovsky,
  A.~Whitcombe, N.~Narevsky, G.~Wright, T.~A. Courtade, E.~Alon, B.~Nikolic,
  and A.~M. Niknejad, ``Design of energy- and cost-efficient {m}assive {MIMO}
  arrays,'' \emph{Proceedings of the IEEE}, vol. 104, no.~3, pp. 586--606, 2016.

\bibitem{li2017decentralized}
K.~Li, R.~R. Sharan, Y.~Chen, T.~Goldstein, J.~R. Cavallaro, and C.~Studer,
  ``Decentralized baseband processing for {Massive MU-MIMO} systems,''
  \emph{IEEE Journal on Emerging and Selected Topics in Circuits and Systems},
  vol.~7, no.~4, pp. 491--507, Dec. 2017.

\bibitem{DBLP:conf/sigcomm/YangLYFTHZZ13} Q.~Yang, X.~Li, H.~Yao,
  J.~Fang, K.~Tan, W.~Hu, J.~Zhang, and Y.~Zhang, ``Bigstation:
  enabling scalable real-time signal processing in large {MU-MIMO}
  systems,'' in \emph{Prof. of ACM SIGCOMM}, pp. 399--410, Aug. 2013.

\bibitem{Bertilsson2017} E.~Bertilsson, O.~Gustafsson, and
  E.~G. Larsson, ``{A scalable architecture for {m}assive {MIMO} base
    stations using distributed processing},'' in \emph{Proc. of
    Asilomar Conference on Signals, Systems and Computers}, pp.
  864--868, Nov. 2016.

\bibitem{Auer2011}
G.~Auer, V.~Giannini, C.~Desset, I.~Godor, P.~Skillermark, M.~Olsson, M.~A.
  Imran, D.~Sabella, M.~J. Gonzalez, O.~Blume, and A.~Fehske, ``How much energy
  is needed to run a wireless network?'' \emph{IEEE Wireless Comm.},
  vol.~18, no.~5, pp. 40--49, Oct. 2011.

\bibitem{DBLP:journals/corr/abs-1712-09612}
\BIBentryALTinterwordspacing
C.~Moll{\'{e}}n, U.~Gustavsson, T.~Eriksson, and E.~G. Larsson, ``Impact of
  spatial filtering on distortion from low-noise amplifiers in massive {MIMO}
  base stations,'' \emph{IEEE Trans. Comm.}, 2018. To appear.

\bibitem{Li2011} Y.~Li, J.~Lopez, P.~H. Wu, W.~Hu, R.~Wu, and
  D.~Y.~C. Lie, ``A {SiGe} envelope-tracking {P}ower {A}mplifier with
  an integrated {CMOS} envelope modulator for mobile {WiMAX/3GPP LTE}
  transmitters,'' \emph{IEEE Trans. on Microwave Theory and
    Techniques}, vol.~59, no.~10, pp. 2525--2536, Oct. 2011.

\bibitem{Horlin2008}
P.~Horlin and A.~Bourdoux, \emph{Digital Compensation for Analog Front-Ends: A
  New Approach to Wireless Transceiver Design}.\hskip 1em plus 0.5em minus
  0.4em\relax Wiley, 2008.

\bibitem{Larsson2017PA} E.~G. Larsson and L.~Van der Perre,
  ``Out-of-band radiation from antenna arrays clarified,'' \emph{IEEE
  Wireless Comm. Lett.}, 2018. To appear.

\bibitem{DBLP:journals/corr/abs-1711-02439}
\BIBentryALTinterwordspacing
C.~Moll{\'{e}}n, U.~Gustavsson, T.~Eriksson, and E.~G. Larsson, ``Spatial
  characteristics of distortion radiated from antenna arrays with transceiver
  nonlinearities,'' \emph{CoRR}, vol. abs/1711.02439, 2017. [Online].
  Available: \url{http://arxiv.org/abs/1711.02439}
\BIBentrySTDinterwordspacing

\bibitem{Pelgrom17book}
M.~Pelgrom, \emph{Analog-to-Digital Conversion}.\hskip 1em plus 0.5em minus
  0.4em\relax Springer, 2017.

\bibitem{mollen1bit}
C.~Moll{\'e}n, J.~Choi, E.~G. Larsson, and J.~R.~W.~Heath, ``Uplink performance
  of wideband {m}assive {MIMO} with one-bit {ADCs},'' \emph{IEEE Trans.
  Wireless Comm.}, vol.~16, pp. 87--100, Jan. 2017.

\bibitem{mollen2016achievable} C.~Moll{\'e}n, J.~Choi, E.~G. Larsson,
  and R.~W. Heath~Jr., ``Achievable uplink rates for {m}assive {MIMO}
  with coarse quantization,'' in \emph{Proc. of IEEE International
    Conference on Acoustics, Speech and Signal Processing (ICASSP)},
 Mar. 2017.

\bibitem{7931630}
Y.~Li, C.~Tao, G.~Seco-Granados, A.~Mezghani, A.~L. Swindlehurst, and L.~Liu,
  ``Channel estimation and performance analysis of one-bit {m}assive {MIMO}
  systems,'' \emph{IEEE Trans. Signal Process.}, vol.~65, no.~15, pp.
  4075--4089, Aug. 2017.

\bibitem{fan2015uplink}
L.~Fan, S.~Jin, C.-K. Wen, and H.~Zhang, ``Uplink achievable rate for {m}assive
  {MIMO} with low-resolution {ADC},'' \emph{IEEE Commun. Lett.}, vol.~19,
  no.~12, pp. 2186--2189, Dec. 2015.

\bibitem{zhang2016spectral}
J.~Zhang, L.~Dai, S.~Sun, and Z.~Wang, ``On the spectral efficiency of
  {m}assive {MIMO} systems with low-resolution {ADCs},'' \emph{IEEE Commun.
  Lett.}, vol.~20, no.~5, pp. 842--845, May 2016.

\bibitem{vanderplas2008}
G.~Van~der Plas and B.~Verbruggen, ``{A 150 MS/s 133 $\mu$W 7 bit ADC in 90nm
  Digital CMOS},'' \emph{IEEE Journal of Solid-State Circuits}, no. 43-12, pp.
  2631--2640, Dec. 2008.

\bibitem{Choo2016} K.~Choo, J.~Bell, and M.~Flynn, ``{Area-efficient
  1GS/s 6b SAR ADC with charge-injection-cell-based DAC},'' in
  \emph{Proc. of IEEE International Solid-State Circuits Conference
    (ISSCC)}, Feb. 2016.

\bibitem{7887699}
Y.~Li, C.~Tao, A.~L. Swindlehurst, A.~Mezghani, and L.~Liu, ``Downlink
  achievable rate analysis in {m}assive {MIMO} systems with one-bit {DACs},''
  \emph{IEEE Commun. Lett.}, vol.~21, no.~7, pp. 1669--1672, Jul. 2017.

\bibitem{Jacobsson2017rate} S.~Jacobsson, G.~Durisi, M.~Coldrey,
  T.~Goldstein, and C.~Studer, ``Quantized precoding for {massive
    MU-MIMO},'' \emph{IEEE Trans.  Comm.}, vol.~65, no.~11,
  pp. 4670--4684, Nov. 2017.

\bibitem{Desset2015} C.~Desset and L.~{Van der Perre}, ``Validation of
  low-accuracy quantization in {m}assive {MIMO} and constellation
  {EVM} analysis,'' in \emph{Proc. of European Conference on Networks
    and Communications (EuCNC)}, Jun. 2015.

\bibitem{Jacobsson2017} S.~Jacobsson, G.~Durisi, M.~Coldrey, and
  C.~Studer, ``On out-of-band emissions of quantized precoding in
  massive {MU-MIMO-OFDM},'' in \emph{Proc. of Asilomar Conference on
    Signals, Systems, and Computers}, Oct. 2017.

\bibitem{MAMMOETD2_4}
\BIBentryALTinterwordspacing
{MAMMOET} project deliverable {D2.4}. [Online]. Available:
  \url{https://mammoet-project.eu/downloads/publications/deliverables/MAMMOET-D2.4-Analysis-non-reciprocity-impact-PU-M20.pdf}
\BIBentrySTDinterwordspacing

\bibitem{Bourdoux2003}
A.~Bourdoux, B.~Come, and N.~Khaled, ``Non-reciprocal transceivers in
  {OFDM/SDMA} systems: impact and mitigation,'' in \emph{Proc. of Radio and Wireless
  Conference (RAWCON)}, pp. 183--186, Aug. 2003.

\bibitem{Vieira2017} J.~Vieira, F.~Rusek, O.~Edfors, S.~Malkowsky,
  L.~Liu, and F.~Tufvesson, ``Reciprocity calibration for {m}assive
  {MIMO}: Proposal, modeling and validation,'' \emph{IEEE Trans. on
    Wireless Comm.}, vol.~16, no.~5, pp. 3042--3056, Mar. 2017.

\bibitem{ZhuICC2015} D.~Zhu, B.~Li, and P.~Liang, ``{On the matrix
  inversion approximation based on Neumann series in {m}assive {MIMO}
  systems},'' in \emph{Proc. of IEEE International Conference on Communications (ICC)}, pp. 1763--1769,
  Jun. 2015.

\bibitem{Mueller2016EURASIP} A.~Mueller, A.~Kammoun, E.~Bj{\"o}rnson,
  and M.~Debbah, ``Linear precoding based on polynomial expansion:
  Reducing complexity in {m}assive {MIMO},'' \emph{EURASIP Journal on
    Wireless Comm. and Netw.}, Dec. 2016.

\bibitem{Hemanth2014ISCAS} H.~Prabhu, O.~Edfors, J.~Rodrigues, L.~Liu,
  and F.~Rusek, ``Hardware efficient approximative matrix inversion
  for linear pre-coding in {m}assive {MIMO},'' in \emph{Proc. of IEEE
    International Symposium on Circuits and Systems (ISCAS)},
  pp. 260--265, Jun. 2014.

\bibitem{Wu2014}
M.~Wu, B.~Yin, G.~Wang, C.~Dick, J.~Cavallaro, and C.~Studer, ``Large-scale
  {MIMO} detection for {3GPP LTE}: Algorithms and {FPGA} implementations,''
  \emph{IEEE Journal of Sel. Topics in Sign. Proc.}, vol.~8, no.~5,
  pp. 916--929, Oct. 2014.

\bibitem{Lee2016tvt}
K.~Lee and C.~Chen, ``An eigen-based approach for enhancing matrix inversion
  approximation in {m}assive {MIMO} systems,'' \emph{IEEE Trans. 
  Veh. Techn.}, vol.~66, no.~6, pp. 5483--5487, Jun. 2017.

\bibitem{Nagy2017WCL}
B.~Nagy, M.~Elsabrouty, and S.~Elramly, ``Fast converging weighted Neumann
  series precoding for {m}assive {MIMO} systems,'' \emph{IEEE Wireless
  Comm. Lett.}, Oct. 2017.

\bibitem{Yin2014}
B.~Yin, M.~Wu, J.~R. Cavallaro, and C.~Studer, ``Conjugate gradient-based
  soft-output detection and precoding in {massive MIMO} systems,'' in
  \emph{Proc. of IEEE Global Communications Conference (GLOBECOM)}, pp. 3696--3701, Dec. 2014.

\bibitem{Wu2016} M.~Wu, C.~Dick, J.~Cavallaro, and C.~Studer,
  ``High-throughput data detection for {m}assive {MU-MIMO-OFDM} using
  coordinate descent,'' \emph{IEEE Trans. Circ. and Syst. I: Regular
    Papers}, vol.~63, no.~12, pp.  2357--2367, Dec. 2016.

\bibitem{Gao2015ICC} X.~Gao, L.~Dai, J.~Zhang, S.~Han, and
  I.~Chih-Lin, ``{Capacity-approaching linear precoding with
    low-complexity for large-scale {MIMO} systems},'' in
  \emph{Proc. of IEEE International Conference on Communications
    (ICC)}, pp. 1577--1582, Jun. 2015.

\bibitem{Prabhu2017} H.~Prabhu, J.~Rodrigues, L.~Liu, and O.~Edfors,
  ``A 60 p{J}/b 300 {M}b/s 128 x 8 {m}assive {MIMO} precoder-detector
  in 28 nm {FD}-{SOI},'' in \emph{Proc. of IEEE International Solid-State
    Circuits Conference (ISSCC)}, Feb. 2017.

\bibitem{Alshamary2016ISIT} H.~A.~J. Alshamary and W.~Xu, ``Efficient
  optimal joint channel estimation and data detection for {m}assive
  {MIMO} systems,'' in \emph{Proc. of IEEE International Symposium on
    Information Theory (ISIT)}, pp. 875--879, Jul. 2016.

\bibitem{Rakesh2017ISCAS} R.~Gangarajaiah, H.~Prabhu, O.~Edfors, and
  L.~Liu, ``A {Cholesky} decomposition based {m}assive {MIMO} uplink
  detector with adaptive interpolation,'' in \emph{Proc. of IEEE
    International Symposium on Circuits and Systems (ISCAS)}, May 2017.

\bibitem{Mirsad2014} M.~Cirkic and E.~Larsson, ``{On the complexity of
  very large multi-user {MIMO} detection},'' in \emph{Proc. of IEEE
  International Workshop on Signal Processing Advances in Wireless
  Communications (SPAWC)}, pp. 55--59, Jun. 2014.

\bibitem{Tang2018} T.~Wei, H.~Prabhu, L.~Liu, {\"O}.~Viktor, and
  Z.~Zhengya, ``A 1.8{G}b/s 70.6p{J}/b 128×16 link-adaptive
  near-optimal massive {MIMO} detector in 28nm {UTBB-FDSOI},'' in
  \emph{Proc. of IEEE International Solid-State Circuits Conference
    (ISSCC)}, Feb. 2018.

\bibitem{Nicolas2012} N.~Planes, O.~Weber, V.~Barral, S.~Haendler,
  D.~Noblet, D.~Croain, M.~Bocat, P.~Sassoulas, X.~Federspiel,
  A.~Cros, and A.~Bajolet, ``28nm {FDSOI} technology platform for
  high-speed low-voltage digital applications,'' in \emph{Proc. of
    IEEE Symposium on VLSI Technology (VLSIT)}, pp. 133--134, Jun. 2012.

\bibitem{Huang2017} Y.~Huang, C.~Desset, A.~Bourdoux, W.~Dehaene, and
  L.~{Van der Perre}, ``Massive {MIMO} processing at the semiconductor
  edge: Exploiting the system and circuit margins for power savings,''
  in \emph{Proc. of IEEE International Conference on Acoustics, Speech
    and Signal Processing (ICASSP)}, pp. 3474--3478, Mar. 2017.

\bibitem{Gunnar2017} S.~Gunnarsson, M.~Bortas, Y.~Huang, C.-M. Chen,
  L.~{Van der Perre}, and O.~Edfors, ``Lousy processing increases
  energy efficiency in {m}assive {MIMO} systems,'' in \emph{Proc. of
    European Conference on Networks and Communications (EuCNC)},
  Jun. 2017.

\bibitem{Huang2016} Y.~Huang, M.~Li, C.~Li, P.~Debacker, and L.~{Van
  der Perre}, ``{Computation-skip error mitigation scheme for power
  supply voltage scaling in recursive applications},'' \emph{Journal
  of Signal Processing Systems}, vol.~84, no.~3, pp. 413--424,
  Sep. 2016.

\bibitem{Han2013} J.~Han and M.~Orshansky, ``{Approximate computing:
  An emerging paradigm for energy-efficient design},'' in
  \emph{Proc. of European Test Symposium (ETS)}, May 2013.

\bibitem{Saha2010} S.~K. Saha, ``Modeling process variability in
  scaled {CMOS} technology,'' \emph{IEEE Design Test of Computers},
  vol.~27, no.~2, pp. 8--16, Mar. 2010.

\bibitem{fojtik2012} M.~Fojtik, D.~Fick, Y.~Kim, N.~Pinckney,
  D.~Harris, D.~Blaauw, and D.~Sylvester, ``{Bubble Razor: An
    architecture-independent approach to timing-error detection and
    correction},'' in \emph{Proc. of IEEE International Solid-State
    Circuits Conference (ISSCC)}, pp. 488--490, Feb. 2012.

\bibitem{liu2010} Y.~Liu, T.~Zhang, and K.~K. Parhi, ``{Computation
  error analysis in digital signal processing systems with overscaled
  supply voltage},'' \emph{IEEE Trans. on Very Large Scale
  Integr. Syst.}, vol.~18, pp. 517--526, Apr. 2010.

\bibitem{ITU_5G} \BIBentryALTinterwordspacing {ITU} vision on {5G}
  usage scenarios. [Online]. Available:
  \url{https://www.itu.int/dms\_pubrec/itu-r/rec/m/R-REC-M.2083-0-201509-I!!PDF-E.pdf}
  \BIBentrySTDinterwordspacing

\bibitem{Vandenameele2000} P.~Vandenameele, L.~{Van der Perre},
  M.~G.~E. Engels, B.~Gyselinckx, and H.~J.~D. Man, ``A combined
  {OFDM/SDMA} approach,'' \emph{IEEE Journal on Sel. Areas in Comm.},
  vol.~18, no.~11, pp. 2312--2321, Nov. 2000.

\bibitem{Yoshizawa2009} Y. Shingo and Y.~Miyanaga, ``{VLSI}
  implementation of a 4x4 {MIMO-OFDM} transceiver with an 80-{MH}z
  channel bandwidth,'' in \emph{Proc. of IEEE International Symposium
    on Circuits and Systems}, pp. 1743--1746, May 2009.

\bibitem{Ketonen2010} J.~Ketonen, M.~Juntti, and J.~R. Cavallaro,
  ``Performance-complexity comparison of receivers for a {LTE
  MIMO-OFDM} system,'' \emph{IEEE Trans. on Sign. Proc.}, vol.~58,
  no.~6, pp. 3360--3372, Jun. 2010.

\bibitem{DRAMCO_tutorial} \BIBentryALTinterwordspacing Connecting
  sensors with low power wireless technologies. [Online]. Available:
  \url{https://dramco.be/tutorials/low-power-lora/}
  \BIBentrySTDinterwordspacing

\bibitem{Malkowsky2017IEEEACESS}
S.~Malkowsky, J.~Vieira, L.~Liu, P.~Harris, K.~Nieman, N.~Kundargi, I.~Wong,
  F.~Tufvesson, V.~Owall, and O.~Edfors, ``The world's first real-time testbed
  for {m}assive {MIMO}: Design, implementation, and validation,'' \emph{IEEE
  Access}, vol.~5, pp. 9073--9088, May 2017.

\bibitem{TestbedEurecom}
``Openairinterface {M}assive {MIMO} testbed : A {5G} innovation platform,''
  \url{http://www.openairinterface.org/}.

\bibitem{TestbedFacebook}
``Introducing facebook's new terrestrial connectivity systems -- terragraph
  and project {ARIES},''
  \url{https://code.facebook.com/posts/1072680049445290/introducing-facebook-s-new-terrestrial-connectivity-systems-terragraph-and-project-aries/}.

\bibitem{Chen2016} C.~M. Chen, V.~Volskiy, A.~Chiumento, {L. Van der
  Perre}, G.~A.~E. Vandenbosch, and S.~Pollin, ``Exploration of user
  separation capabilities by distributed large antenna arrays,'' in
  \emph{Proc. of IEEE Global Communications Conference (GLOBECOM)
    Workshops}, Dec. 2016.

\bibitem{Harris2016SIPS} P.~Harris, W.~Hasan, S.~Malkowsky, J.~Vieira,
  S.~Zhang, M.~Beach, L.~Liu, E.~Mellios, A.~Nix, S.~Armour, and
  A.~Doufexi, ``Serving 22 users in real-time with a 128-antenna
  {m}assive {MIMO} testbed,'' in \emph{Proc. of IEEE International
    Workshop on Signal Processing Systems (SiPS)}, pp.  266--272,
  Oct. 2016.

\bibitem{MAMMOETD4_2}
\BIBentryALTinterwordspacing
{MAMMOET} project deliverable {D4.2}. [Online]. Available:
  \url{https://mammoet-project.eu/downloads/publications/deliverables/MAMMOET-D4.2-Testbed-assessment-PU-M33.pdf}
\BIBentrySTDinterwordspacing

\bibitem{MaMiBlog2016} ``Massive {MIMO} in mobile environments,'' in
  the \emph{Massive MIMO blog}, \url{http://massive-mimo.net}.

\bibitem{Barati2015TWC} C.~Barati, S.~Hosseini, S.~Rangan, P.~Liu,
  T.~Korakis, S.~Panwar, and T.~Rappaport, ``Directional cell
  discovery in millimeter wave cellular networks,'' \emph{IEEE
    Trans. on Wireless Comm.}, vol.~14, no.~2, pp. 6664--6678,
  Dec. 2015.

\bibitem{Karlsson2017arXiv} M.~Karlsson, E.~Bj{\"o}rnson, and
  E.~G. Larsson, ``Performance of in-band transmission of system
  information in {m}assive {MIMO} systems,'' \emph{IEEE
    Trans. Wireless Commun.}, vol.~17, pp.~1700--1712, Mar.~2018.

\bibitem{Xia16} X.~G. Xia and X.~Gao, ``A space-time code design for
  omnidirectional transmission in {m}assive {{MIMO}} systems,''
  \emph{IEEE Wireless Comm. Lett.}, vol. 5, no. 5, pp. 512--515,
  Oct. 2016.

\bibitem{Meng16} X.~Meng, X.~Gao, and X.~G. Xia, ``Omnidirectional
  precoding based transmission in {m}assive {{MIMO}} systems,''
  \emph{IEEE Trans. on Comm.}, vol.~64, no.~1, pp. 174--186,
  Jan. 2016.

\bibitem{Marco2016CM}
G.~Marco, M.~Mezzavilla, and M.~Zorzi, ``Initial access in 5{G} mmwave cellular
  networks,'' \emph{IEEE Comm. Mag.}, vol.~54, no.~11, pp. 40--47,
  Nov. 2016.

\bibitem{FranceschettiM2015} M.~Franceschetti, ``On {L}andau's
  eigenvalue theorem and information cut-sets,'' \emph{IEEE
    Trans. Inf. Theory}, vol.~61, no.~9, pp. 5042--5051, Sep. 2015.

\bibitem{ngo2017cell} H.~Q. Ngo, A.~Ashikhmin, H.~Yang, E.~G. Larsson,
  and T.~L. Marzetta, ``Cell-free massive {MIMO} versus small cells,''
  \emph{IEEE Trans. Wireless Commun.}, vol.~16, no.~3, pp. 1834--1850,
  Mar. 2017.

\bibitem{LIS2017} S.~Hu, F.~Rusek, and O.~Edfors, ``Beyond {massive
  MIMO}: The potential of data transmission with large intelligent
  surfaces,'' \emph{IEEE Trans. on Sign. Proc.}, vol.~66, no.~10,
  pp. 2746--2758, May 2018.

\end{thebibliography}
